# Fisher Lecture: Dimension Reduction in Regression[1,2]

R. Dennis Cook

*Abstract.* Beginning with a discussion of R. A. Fisher's early written remarks that relate to dimension reduction, this article revisits principal components as a reductive method in regression, develops several model-based extensions and ends with descriptions of general approaches to model-based and model-free dimension reduction in regression. It is argued that the role for principal components and related methodology may be broader than previously seen and that the common practice of conditioning on observed values of the predictors may unnecessarily limit the choice of regression methodology.

## 1. INTRODUCTION

R. A. Fisher is responsible for the context and mathematical foundations of a substantial portion of contemporary theoretical and applied statistics. One purpose of this article is to consider insights into the long-standing and currently prominent problem of "dimension reduction" that may be available from his writings. Two papers are discussed in this regard, Fisher's pathbreaking 1922 paper on the theoretical foundations of statistics (Section 1.1), and a later applications paper on the yield of wheat at Rothamsted (Section 1.2). The discussion of these articles makes liberal use of quoted material, in an effort to preserve historical flavor and reflect Fisher's style.

Principal component analysis is one of the oldest and best known methods for reducing dimensionality in multivariate problems. Another purpose of this article is to provide an exposition on principal components as a reductive method in regression, with emphasis on connections to known reductive methods and on the development of a new method—*principal fitted components* ($PFC$)—that may outperform principal components. The discussion will be related to and guided by Fisher's writings as much as the nature of the case permits. In particular, the philosophical spirit of the methods discussed in this article derives largely from Fisher's notion of sufficiency.

We briefly review principal components in Section 2, and starting in Section 3 we focus on principal components in regression. Principal fitted components are introduced in Section 4. In Sections 5–7 we expand the themes of Sections 3 and 4, gradually increasing the scope of dimension reduction methodology. A general model-based paradigm for dimension reduction in regression is described in Section 8.1. In keeping with the Fisherian theme of this article, the development is model-based, but in

*R. Dennis Cook is Professor, School of Statistics, University of Minnesota, 224 Church Street S. E., Minneapolis, Minnesota 55455, USA e-mail: dennis@stat.umn.edu.*

[1]This article is the written version of the 2005 Fisher lecture delivered at the Joint Statistical Meetings in Minneapolis, Minnesota.

[2]Discussed in 10.1214/088342307000000041, 10.1214/088342307000000050 and 10.1214/088342307000000069; rejoinder at 10.1214/088342307000000078.

Section 8.2 we describe its relation to recent ideas for model-free reductions. The Appendix contains justification for propositions and other results. Emphasis is placed on ideas and methodological directions, rather than on the presentation of fully developed methods.

Reduction by principal components has been proposed as adjunct methodology for linear regression. It does not arise as a particular consequence of the model itself but is used to mitigate the variance inflation that often accompanies collinearities among the predictors. Indeed, while collinearity is the main and often the only motivation for use of principal components in regression, it will play no role in the evolution of the methods in this article. It is argued in Sections 5.1, 5.2 and 8 that the utility of principal component reduction is broader than previously seen and need not be tied to the presence of collinearity. This conclusion is a consequence of postulating inverse regression models that lead to principal components and principal fitted components as maximum likelihood estimators of reductive subspaces.

Ordinary least squares (OLS) is widely recognized as a reasonable first method of regression when the response and predictors follow a nonsingular multivariate normal distribution. Nevertheless, examples are given in Sections 5 and 6.4 to demonstrate that in this context reduction by principal components and principal fitted components may dominate OLS without invoking collinearity. Sliced inverse regression (SIR; Li, 1991) is a relatively recent reductive method for regression that has received notable attention in the literature. New drawbacks of SIR will emerge from its relation to principal fitted components described in Sections 6.3 and 7.5. It is demonstrated by example that the method of principal fitted components can dominate SIR as well as OLS in the multivariate normal setting. The notions of principal components and principal fitted components are presented here as reductive frames not necessarily tied to normality. Their construction in the context of exponential families is sketched in Section 3.3 and elsewhere.

Conditioning on the observed values of the predictors is a well-established practice in regression, even when the response and the predictors have a joint distribution. However, as a consequence of the exposition on principal components and related reductive methods, we argue in Section 8 that this practice may unnecessarily restrict our choice of regression methodology. It may be advantageous in some regressions to make explicit use of the variability in the predictors through their multivariate inverse regression on the response.

### 1.1 Fisher, 1922

Much of contemporary statistical thought began with Fisher's 1922 article "On the mathematical foundations of theoretical statistics," which set forth a new conceptual framework for statistics, including definitions of Consistency, Efficiency, Likelihood, Specification and Sufficiency. Fisher also introduced the now familiar terms "statistic," "maximum likelihood estimate" and "parameter." The origins of this remarkable work, which in many ways is responsible for the ambient texture of statistics today, were traced by Stigler (1973, 2005), who emphasized that Fisher was the first to use the word "parameter" in its modern context, 57 times in his 1922 article (Stigler, 1976). More than any other single notion, this word reflects the starting point—parametric families—for Fisher's constructions. Many of his ideas are now common knowledge, to the point that he is no longer credited when they are first introduced in some statistics texts. This paper, perhaps more than any other of Fisher's, should be required reading in every statistics curriculum.

Fisher provided a focal point for statistical methods at the outset of his 1922 article:

> ...the objective of statistical methods is the reduction of data. A quantity of data... is to be replaced by relatively few quantities which shall adequately represent... the relevant information contained in the original data.
>
> Since the number of independent facts supplied in the data is usually far greater than the number of facts sought, much of the information supplied by an actual sample is irrelevant. It is the object of the statistical process employed in the reduction of data to exclude this irrelevant information, and to isolate the whole of the relevant information contained in the data.

In these statements Fisher signified the goal of statistical methods as a type of dimension reduction, the reduced data containing the relevant and only the relevant information. The fundamental idea that statistical methods deal with the reduction of data did not originate with Fisher, and was probably widely understood well before his birth in 1890 (see, e.g., Edgeworth, 1884). However, Fisher's approach is unique because it changed the course of statistical history.

Fisher identified three distinct problems in his reductive process: 1. "Problems of Specification," selecting a parametric family; 2. "Problems of Estimation," selecting statistics for parameter estimation; and 3. "Problems of Distribution," deriving the sampling distribution of the selected statistics or functions thereof. Problems of specification and estimation are both reductive in nature. Surely, selecting a finitely and parsimoniously parameterized family from the infinity of possible choices is a crucial first reductive step that can overshadow any subsequent reduction of the data for the purpose of estimation.

Regarding estimation, Fisher said that once the model is specified

> ...the statistic chosen should summarize the whole of the relevant information supplied by the sample.

Any operational version of this idea must include some way of parsing data into the relevant and irrelevant. For Fisher, the reductive process started with a parametric family, targeted information on its parameters $\theta$ and was guided by sufficiency: If $D$ represents the data, then a statistic $t(D)$ is sufficient if

$$(1) \qquad D|(\theta, t) \sim D|t$$

so that $t$ contains all of the relevant information about $\theta$. Subsequent commentaries on sufficiency by Fisher and others address existence, minimal sufficiency, specializations and variations, relation to the method of maximum likelihood, and the factorization theorem.

Fisher was quite specific on how to approach the second reductive step, estimation, but was less so regarding the overarching first reductive step, specification. In stating that problems of specification "...are entirely a matter for the practical statistician," Fisher positioned them as a nexus between applied and theoretical statistics. A model is required before proceeding to problems of estimation, and model specification falls under the purview of the practical statistician. He also offered the following helpful but nonprescriptive advice:

> ...we may know by experience what forms are likely to be suitable, and the adequacy of our choices may be tested *a posteriori*. We must confine ourselves to those forms which we know how to handle...."

In these statements Fisher acknowledged a role for statisticians as members of scientific teams, anticipated the development of diagnostic methods for model criticism and recognized a place for off-the-shelf models. Interest in diagnostic methods for regression was particularly high from a period starting near the time of Fisher's death in 1962 and ending in the late 1980s (Anscombe, 1961; Anscombe and Tukey, 1963; Box, 1980; Cook and Weisberg, 1982; Cook, 1986). Fisher also linked model complexity with the amount of data, and evidently did not require that models be "true,"

> More or less elaborate forms will be suitable according to the volume of the data. Evidently these are considerations the nature of which may change greatly during the course of a single generation.

Fisher's first reductive step is the most challenging and elusive. Indeed, Fisher's views launched a debate over modeling that continues today. Writing on Fisher's discovery of sufficiency, Stigler's (1973) introduction began with the sentence:

> Because Fisher's concept of sufficiency depends so strongly on the assumed form of the population distribution, its importance to applied statistics has been questioned in recent years.

Box's (1979) memorable statement that "All models are wrong, but some are useful" has been taken to imply, perhaps from a position of devil's advocate, that $D$ is the only sufficient statistic. At least two Fisher lectures, Lehmann in 1988 (Lehmann, 1990) and Cox in 1989 (Cox, 1990), were on the issue of model specification. And then there are the modeling cultures of Breiman (2001) and McCullagh's (2002) rather esoteric answer to the question "What is a statistical model?" Fourteen years after his 1922 article, Fisher suggested that it might be possible to develop inductive arguments for model specification, a promise that has yet to be realized:

> Clearly, there can be no operation properly termed 'estimation' until the parameter to be estimated has been well defined, and this requires that the mathematical form of the distribution shall be given. Nevertheless, we need not close our eyes to the possibility that an even wider type of inductive argument may some day be developed, which shall discuss methods of assigning from the data the functional form of the population (Fisher, 1936).

## 1.2 Fisher, 1924

In Fisher's classic 1922 article we see him thinking primarily as a theoretical statistician, while in his 1924 paper "III. The influence of rainfall on the yield of wheat at Rothamsted" we see him as an applied statistician. Having recently developed a new framework for theoretical statistics, he might have quickly integrated those ideas into his applied work, but that does not appear to be the case. He did rely on his 1922 article to justify a claim that a skewness statistic is the "most efficient statistic" for a test of normality (Fisher, 1924, page 103), but otherwise I found no clear formal links between the two works.

The opening issue that Fisher addressed in 1924 involves sparsity of data in regression, possibly "$n < p$." According to Fisher, large sample regression methods are appropriate when the sample size $n$ is much larger than the number of predictors $p$, preferably $n > 1000$ but at least in the hundreds. However, the number $p$ of meteorological variables that might plausibly affect yield could easily exceed the length $n$ of the longest run of available crop records. Fisher then faced a dimension reduction problem rather like those encountered in the analysis of contemporary genomics data. He did not provide a general solution, but did give a clear opinion on what not to do. It was common practice at the time to preprocess the potential meteorological predictors by plotting them individually against yield to select the predictors for the subsequent regression. Fisher was critical of this practice, concluding that

> The meteorological variables to be employed must be chosen without reference to the actual crop record.

He also reiterated a theme of his 1922 paper, "Relationships of a complicated character should be sought only when long series of crop data are available."

The bulk of Fisher's article is devoted to the development of models for the regression of yield on rainfall, models that were painstakingly tailored to the substantive characteristics of the problem, as one would expect in careful application. Fisher's study seems true to his general point that model specification is "entirely a matter for the practical statistician," and the corollary that it depends strongly on context. Regarding dimension reduction, there is at least one general theme in Fisher's analysis: An "$n < p$" regression might be usefully transformed into an "$n > p^*$" regression, but the methodology for doing so should not depend on the response.

Sufficiency aside (and that is a lot to set aside), I found little transportable methodology in Fisher's writing that might be applied to contemporary dimension reduction problems in regression. Nevertheless, dimension reduction was an issue during Fisher's era and before. In the next section we turn to a widely recognized dimension reduction method in statistics—principal components—whose beginnings occurred well before Fisher's time.

## 2. INTRODUCTION TO PRINCIPAL COMPONENTS

It is well established that principal components are a useful and important foundation for reducing the dimension of a multivariate random sample, represented here by the vectors $\mathbf{X}_1, \mathbf{X}_2, \ldots, \mathbf{X}_n$ in $\mathbb{R}^p$. Letting $\lambda_1 \geq \cdots \geq \lambda_p$ and $\boldsymbol{\gamma}_1, \ldots, \boldsymbol{\gamma}_p$ denote the eigenvalues and corresponding vectors of $\boldsymbol{\Sigma} = \text{Var}(\mathbf{X})$, the population principal components are defined as the linearly transformed variables $\{\boldsymbol{\gamma}_1^T \mathbf{X}, \ldots, \boldsymbol{\gamma}_p^T \mathbf{X}\}$. We call the $\boldsymbol{\gamma}_j$'s the *principal component (PC) directions*. The sample principal components are $\{\hat{\boldsymbol{\gamma}}_1^T \mathbf{X}, \ldots, \hat{\boldsymbol{\gamma}}_p^T \mathbf{X}\}$, where $\hat{\boldsymbol{\gamma}}_1, \ldots, \hat{\boldsymbol{\gamma}}_p$ are the eigenvectors (sample PC directions) corresponding to eigenvalues $\hat{\lambda}_1 > \cdots > \hat{\lambda}_p$ of the usual sample covariance matrix $\widehat{\boldsymbol{\Sigma}}$.

The history of principal components goes back at least to Adcock (1878) who wished to

> Find the most probable position of the straight line determined by the measured coordinates, each measure being equally good or of equal weight, $(x_1, y_1), (x_2, y_2), \ldots, (x_n, y_n)$ of $n$ points....

Adcock identified his solution as the "principal axis," or first sample PC direction. Subsequent milestones were given in articles by Pearson (1901), Spearman (1904) and Hotelling (1933), but I found no direct reference to principal components in Fisher's writings. It has been discovered over time that the leading principal components, say $\{\boldsymbol{\gamma}_1^T \mathbf{X}, \ldots, \boldsymbol{\gamma}_k^T \mathbf{X}\}$, $k \ll p$, have a number of properties that may be helpful, depending on application-specific requirements.

Reviews of principal components were given by Seber (1984), Christensen (2001) and Jolliffe (2002). Gould (1981, Chapter 6) provided an interesting and illuminating historical account on the use of principal components in the social sciences. New properties and methods seem to be communicated often in contemporary statistical literature (see, e.g., Jong and Kotz, 1999; Tipping and Bishop, 1999; Maronna, 2005).

Principal components have also been studied in the context of regression, where $\mathbf{X}$ is now the vector of predictors that we would like to reduce prior to performing a regression with response $Y$. There are several reasons why such reduction may be useful in practice, including the possibilities of mitigating the effects of collinearity, facilitating model specification by allowing visualization of the regression in low dimensions (Cook, 1998) and providing a relatively small set of predictors on which to base prediction or interpretation. Collinearity in particular has been the main motivation for using principal components as a reductive method in regression. One persistent idea is that perhaps we can use the leading principal components in place of $\mathbf{X}$ with little loss of information: The first few principal components should contain essentially the same information about $Y$ as the original predictors, which is in the spirit of Fisher's idea of sufficiency. Kendall (1957, page 75), Hocking (1976) and others suggested using principal components in this way. Scott (1992) suggested that the predictors might be "compressed" by using their principal components prior to additive nonparametric modeling. Such procedures have been questioned because principal components are computed from the marginal distribution of $\mathbf{X}$ and consequently the leading components may have little necessary relation with the response. It does seem optimistic to think that the marginal of $\mathbf{X}$ would necessarily be structured so that the leading principal components contain the essential information about the response. Nevertheless, in support of the leading principal components, Mosteller and Tukey (1977, page 397) turned this cause for concern into a desirable goal by posing the question:

> ...how can we find linear combinations of the [predictors] that will be likely, or unlikely, to pick up regression from some as yet unspecified $y$?

It might seem unusual that Mosteller and Tukey would begin by asking how to perform linear reduction without reference to the response, but their approach is in agreement with Fisher's (1924) point that predictors "...be chosen without reference to the actual crop record." Mosteller and Tukey answered their question with the philosophical point that:

> A malicious person who knew our $x$'s and our plan for them could always invent a $y$ to make our choices look horrible. But we don't believe nature works that way— more nearly that nature is, as Einstein put it (in German), "tricky, but not downright mean."

On the other hand, Cox (1968, page 272) wrote in reference to reducing $\mathbf{X}$ by using the leading principal components:

> A difficulty seems to be that there is no logical reason why the dependent variable should not be closely tied to the least important principal component.

A similar sentiment was expressed by Hotelling (1957), Hawkins and Fatti (1984) and others. Evidently, many authors did not trust nature in the same way as Mosteller and Tukey. Some gave examples of regressions where Cox's prediction apparently holds, with a few trailing principal components contributing important information about the response (Jolliffe, 1982; Hadi and Ling, 1998). Are there limits to the maliciousness of nature? If nature can produce regressions where there is useful information about the response in the last principal component $\gamma_p^T \mathbf{X}$, can it also produce settings where the regression information is concentrated in a few of the original predictors, but is spread evenly across many of the principal components, making analysis on the principal component scale harder than analysis in the original scale? This possibility is relevant in recent proposals to ignore the hierarchical structure of the principal components and use instead a general subset $M$ of them when developing linear regression models (Hwang and Nettleton, 2003). In this regard, Jolliffe (2002, page 177) commented that "...the choice of $M$ for PC regression remains an open question."

On balance, the role for principal components in regression seems less clear-cut than their role in reducing $\mathbf{X}$ marginally. The advantages that principal components enjoy in the multivariate setting, where the marginal distribution of $\mathbf{X}$ is of primary interest, may be of limited relevance in the regression context, where the conditional distribution of $Y|\mathbf{X}$ is of interest. It seems particularly appropriate to question the usefulness of principal components when $\mathbf{X}$ is fixed and controlled by the experimenter. In some experimental designs $\widehat{\boldsymbol{\Sigma}}$ is proportional to the identity, with the consequence that there is clearly no useful relation between its eigenstructure and the regression. And yet interest in using principal components for dimension reduction in regression has persisted, notably in the analysis of microarrays where principal components have been called "eigengenes" (Alter, Brown and Botstein, 2000). Chiaromonte and Martinelli (2002) and L. Li and H. Li (2004) used principal components for preprocessing microarray data, allowing subsequent application of other dimension reduction methodology that requires $n > p$. Acknowledging that preprocessing by principal components might be a viable alternative, Bura and Pfeiffer (2003) used marginal $t$-tests from the regression of the response on each predictor for prior selection of genes, a method to which Fisher would likely have objected.

## 3. PRINCIPAL COMPONENTS IN REGRESSION

In the rest of this article we consider only regressions where $\mathbf{X}$ is random, so $Y$ and $\mathbf{X}$ have a joint distribution, since a different structure seems necessary when $\mathbf{X}$ is fixed and subject to experimental control. In this context we may pursue dimension reduction through the conditional distribution of $Y|\mathbf{X}$ (forward regression) the conditional distribution of $\mathbf{X}|Y$ (inverse regression), or the joint distribution of $(Y, \mathbf{X})$. All three settings have been used in arguments for the relevance of principal components and related methodology. Forward regression with fixed predictors is perhaps the most common, particularly in early articles. Helland (1992) and Helland and Almøy (1994) assumed that $(Y, \mathbf{X})$ follows a multivariate normal in their development of "relevant components." Oman (1991) used an inverse model for $\mathbf{X}|Y$ in a heuristic argument that the coefficient vector $\boldsymbol{\alpha}$ in a linear regression model for $Y|\mathbf{X}$ should fall in or close to the space spanned by the first few principal components. Similarly, in their development of multiple-shrinkage principal component regression, George and Oman (1996) used a model for the joint distribution of $(Y, \mathbf{X})$ to reach the same conclusion.

In this article, we concentrate on $\mathbf{X}|Y$, although the goal is still to reduce the dimension of $\mathbf{X}$ with little or no loss of information on $Y|\mathbf{X}$. Model-based forward regression analyses traditionally condition on the observed values of the predictors, a characteristically Fisherian operation, even if $\mathbf{X}$ is random (see Savage, 1976, page 468, and Aldrich, 2005, for further discussion of this point). Nevertheless, the conditional distribution of $\mathbf{X}|Y$ may provide a better handle on reductive information since it can be linked usefully to $Y|\mathbf{X}$ (Proposition 1), and I found nothing in Fisher's writings that would compel consideration of only forward regressions. To facilitate the exposition, let $\mathbf{X}_y$ denote a random variable distributed as $\mathbf{X}|(Y = y)$. Subspaces are indicated as $\mathcal{S}_{(\cdot)}$, where the argument is a matrix whose columns span the subspace, and $\mathbf{U} \perp\!\!\!\perp \mathbf{V}|\mathbf{W}$ means that the random vectors $\mathbf{U}$ and $\mathbf{V}$ are conditionally independent given any value for the random vector $\mathbf{W}$. The notation $\mathbb{R}^{a \times b}$ stands for the space of real matrices of dimension $a \times b$, and $\mathbb{R}^a$ means the space of real vectors of length $a$.

### 3.1 A First Regression Model Implicating Principal Components

Consider the following multivariate model for the inverse regression of $\mathbf{X}$ on $Y$:

$$(2) \qquad \mathbf{X}_y = \boldsymbol{\mu} + \boldsymbol{\Gamma}\boldsymbol{\nu}_y + \sigma\boldsymbol{\varepsilon}.$$

Here $\boldsymbol{\mu} \in \mathbb{R}^p$, $\boldsymbol{\Gamma} \in \mathbb{R}^{p \times d}$, $d < p$, $\boldsymbol{\Gamma}^T\boldsymbol{\Gamma} = I_d$, $\sigma \geq 0$ and $d$ is assumed to be known. To emphasize the conditional nature of the model, $y$ is used to index observations in place of the more traditional "$i$" notation. The coordinate vector $\boldsymbol{\nu}_y \in \mathbb{R}^d$ is an unknown function of $y$ that is assumed to have a positive definite sample covariance matrix and is centered to have mean 0, $\sum_y \boldsymbol{\nu}_y = 0$, but is otherwise unconstrained. The centering of $\boldsymbol{\nu}_y$ in the sample is for later convenience and is not essential. The error vector $\boldsymbol{\varepsilon} \in \mathbb{R}^p$ is assumed to be independent of $Y$, and to be normally distributed with mean 0 and identity covariance matrix. This model specifies that, after translation by the intercept $\boldsymbol{\mu}$, the conditional means fall in the $d$-dimensional subspace $\mathcal{S}_{\boldsymbol{\Gamma}}$ spanned by the columns of $\boldsymbol{\Gamma}$. The vector $\boldsymbol{\nu}_y$ contains the coordinates of the translated conditional mean $\mathrm{E}(\mathbf{X}_y) - \boldsymbol{\mu}$ relative to the basis $\boldsymbol{\Gamma}$. The mean function is quite flexible, even when $d$ is small, say at most 3 or 4. Aside from the subscript $y$, nothing on the right-hand side of this model is observable.

While the mean function for model (2) is permissive, the variance function is restrictive, requiring essentially that the predictors be in the same scale and that, conditional on $Y$, they be independent with the same variance. Nevertheless, this model provides a useful starting point for our consideration of principal components, and may be appropriate for some applications dealing with measurement error, image recognition, microarray data and calibration (Oman, 1991). And it may serve as a useful first model when the data are not plentiful, following the spirit of Fisher's recommendations.

Model (2) is similar in form to the functional model for multivariate problems studied by Anderson (1984), but there are important conceptual differences. Model (2) is a formal statement about the regression of $\mathbf{X}$ on $Y$, with the $\boldsymbol{\nu}_y$'s being fixed because we condition on $Y$, in the same way forward regression models are conditioned on $\mathbf{X}$. Functional models for multivariate data postulate latent fixed effects without conditioning. Model (2) also resembles the traditional model for factor analysis, but again there are important differences. There is no response in a factor analysis model and no conditioning. Instead $\boldsymbol{\nu}$, the vector of common factors, is assumed to be jointly distributed with the error $\varepsilon$, with conditions imposed on the joint distribution to ensure identifiability. Additionally, rotation to obtain a "meaningful" estimate of $\boldsymbol{\Gamma}$ is often of interest in factor analysis, while in this article interest in $\boldsymbol{\Gamma}$ does not extend beyond $\mathcal{S}_{\boldsymbol{\Gamma}}$. Model (2) will lead in directions that are not available when considering functional or factor-analytic models for multivariate data.

The following proposition connects the inverse regression model (2) with the forward regression of $Y$ on $\mathbf{X}$.

PROPOSITION 1. *Under the inverse model* (2), *the distribution of* $Y|\mathbf{X}$ *is the same as the distribution of* $Y|\boldsymbol{\Gamma}^T\mathbf{X}$ *for all values of* $\mathbf{X}$.

According to this proposition, $\mathbf{X}$ can be replaced by the *sufficient reduction* $\boldsymbol{\Gamma}^T\mathbf{X}$ without loss of information on the regression of $Y$ on $\mathbf{X}$, and without specifying the marginal distribution of $Y$ or the conditional distribution of $Y|\mathbf{X}$. Here "sufficient reduction" is used in the same spirit as Fisher's "sufficient statistic." One difference is that sufficient statistics are observable, while sufficient reductions may contain unknown parameters, $\boldsymbol{\Gamma}$ in this instance, and thus need to be estimated. A reasonable estimate of $\boldsymbol{\Gamma}$ is needed for the sufficient reduction $\boldsymbol{\Gamma}^T\mathbf{X}$ to be useful in practice.

### 3.2 Estimation via the Method of Maximum Likelihood

Even with the previously imposed constraint that $\boldsymbol{\Gamma}^T\boldsymbol{\Gamma} = I_d$, $\boldsymbol{\Gamma}$ and $\boldsymbol{\nu}_y$ are not simultaneously estimable under model (2) since, for any orthogonal matrix $\mathbf{O} \in \mathbb{R}^{d \times d}$, we can always rewrite $\boldsymbol{\Gamma}\boldsymbol{\nu}_y = (\boldsymbol{\Gamma}\mathbf{O})(\mathbf{O}^T\boldsymbol{\nu}_y)$, leading to a different factorization. However, the *reductive subspace* $\mathcal{S}_{\boldsymbol{\Gamma}} = \text{span}(\boldsymbol{\Gamma})$ is estimable, and that is the focus of our inquiry. The matrix $\boldsymbol{\Gamma}$ is of interest only by virtue of the subspace generated by its columns, the condition $\boldsymbol{\Gamma}^T\boldsymbol{\Gamma} = I_d$ being imposed for convenience. This implies that two reductions $\boldsymbol{\Gamma}^T\mathbf{X}$ and $\mathbf{A}\boldsymbol{\Gamma}^T\mathbf{X}$ that are connected by a full-rank linear transformation $\mathbf{A}$ are regarded as equivalent for present purposes. To emphasize this distinction we will refer to the parameter space for $\mathcal{S}_{\boldsymbol{\Gamma}}$ as a Grassmann manifold $\mathcal{G}_{p \times d}$ of dimension $d$ in $\mathbb{R}^p$.

Consider a function $G(\mathbf{A})$ that is defined on the set of $p \times d$ matrices with $d < p$ and $\mathbf{A}^T\mathbf{A} = I_d$ and has the property that $G(\mathbf{AO}) = G(\mathbf{A})$ for any orthogonal matrix $\mathbf{O} \in \mathbb{R}^{d \times d}$. The function $G(\mathbf{A})$ depends only on the span of its argument: If $\text{span}(\mathbf{A}) = \text{span}(\mathbf{B})$, then $G(\mathbf{A}) = G(\mathbf{B})$. The set of $d$-dimensional subspaces of $\mathbb{R}^p$ is called a Grassmann manifold, a single point in a Grassmann manifold being a subspace. A Grassmann manifold is the natural parameter space for the $\boldsymbol{\Gamma}$ parameterization in this article. While no technical use will be made of Grassmann manifolds, the terminology is proper in this context and it may serve as a reminder that only the subspace $\mathcal{S}_{\boldsymbol{\Gamma}}$ is of interest. For background on Grassmann manifolds, see Edelman, Arias and Smith (1998) and Chikuse (2003).

Estimation of a sufficient reduction might be based on the method of moments, Bayesian considerations or a concern for robustness, but staying with Fisher we will use the method of maximum likelihood. Assuming that the data consist of a random sample of size $n$ from $(Y, \mathbf{X})$, the maximum likelihood estimator of $\mathcal{S}_{\boldsymbol{\Gamma}}$ can be constructed by holding $\boldsymbol{\Gamma}$ and $\sigma^2$ fixed at possible values $\mathbf{G}$ and $s^2$ and then maximizing the log likelihood over $\boldsymbol{\mu}$ and $\boldsymbol{\nu}_y$. This yields $\hat{\boldsymbol{\mu}} = \bar{\mathbf{X}}$ and, in the absence of replicate $y$'s, a separate value for each of the $n$ $d$-dimensional vectors $\boldsymbol{\nu}_y$ as a function of $(\mathbf{G}, s^2)$: $\boldsymbol{\nu}_y(\mathbf{G}, s^2) = \mathbf{G}^T(\mathbf{X}_y - \bar{\mathbf{X}})$. Substituting back, the partially maximized log likelihood $M_{\text{PC}}$ is then, apart from constants,

$$M_{\text{PC}}(\mathbf{G}, s^2)$$
$$= -(np/2)\log(s^2)$$

$$-(1/2s^2)\sum_y \|\mathbf{X}_y - \bar{\mathbf{X}} - P_{\mathbf{G}}(\mathbf{X}_y - \bar{\mathbf{X}})\|^2$$

$$= -(np/2)\log(s^2) - (n/2s^2)\operatorname{trace}(\widehat{\mathbf{\Sigma}}Q_{\mathbf{G}}),$$

where $P_{\mathbf{G}} = \mathbf{G}\mathbf{G}^T$ is the projection onto $\mathcal{S}_{\mathbf{G}}$ and $Q_{\mathbf{G}} = I_p - P_{\mathbf{G}}$. Although in this and later likelihood functions we use $\mathbf{G}$ as an argument, the function itself depends only on $\mathcal{S}_{\mathbf{G}}$ and thus maximization is over the Grassmann manifold $\mathcal{G}_{p \times d}$. Recalling that $\{\hat{\lambda}_j\}$ and $\{\hat{\boldsymbol{\gamma}}_j\}$ are the eigenvalues and eigenvectors of $\widehat{\mathbf{\Sigma}}$, it follows that the maximum likelihood estimator $\widehat{\mathcal{S}}_{\boldsymbol{\Gamma}}$ of $\mathcal{S}_{\boldsymbol{\Gamma}}$ is

$$\widehat{\mathcal{S}}_{\boldsymbol{\Gamma}} = \operatorname{span}\{\hat{\boldsymbol{\gamma}}_1, \ldots, \hat{\boldsymbol{\gamma}}_d\},$$

and that the estimator of $\sigma^2$ is $\hat{\sigma}^2 = \sum_{j=d+1}^p \hat{\lambda}_j/p$. A sufficient reduction is thus estimated by the first $d$ sample principal components, and for this reason we will refer to model (2) as a *PC regression model*. If there is replication in the observed $y$'s, as may happen if $Y$ is supported on a finite number of points, then the principal components are to be computed from the usual between-class covariance matrix.

In the absence of replication in the $y$'s, the observed responses play no role other than acting collectively as a conditioning argument. The same reduction would be obtained with a continuous multivariate response. In fact, it is not necessary for the response to be observed and thus model (2) is one possible route to satisfying Mosteller and Tukey's desire to reduce $\mathbf{X}$ for an "...as yet unspecified $y$," and is in accord with Fisher's requirement that "...the variables to be employed must be chosen without reference to the actual crop record." However, model (2) does not hold for all potential responses, but requires that the predictors be conditionally independent with common variance.

The fact that all full-rank linear transformations $\mathbf{A}\widehat{\boldsymbol{\Gamma}}^T\mathbf{X}$ of the first $d$ sample principal components $\widehat{\boldsymbol{\Gamma}}^T\mathbf{X}$ are equivalent reductions in the context of model (2) may cast doubt on the usefulness of *post hoc* reification of principal components. The safest interpretations are perhaps the ones that suggest regression mechanisms that can be verified independently of the principal components themselves.

As mentioned previously, the PC model (2) may be appropriate for some applications, but there is also value in the paradigm it suggests for extension to other settings. In the next section we sketch at the conceptual level how the ideas behind (2) can be applied to exponential families, returning to normal models in Section 4.

### 3.3 Extensions to Exponential Families

As in the PC model, we assume that the predictors are independent given $Y$, but instead of requiring $\mathbf{X}_y$ to be normally distributed we assume that the $j$th conditional predictor $X_{yj}$ is distributed according to a one-parameter exponential family with density or mass function of the form

(3) $\quad f_j(x|\eta_{yj}, Y = y) = a_j(\eta_{yj})b_j(x)\exp(x\eta_{yj}).$

We also assume that the natural parameter $\eta_{yj}$ follows the model for the conditional mean $E(\mathbf{X}_y)$ in the normal case,

$$\eta_{yj} = \mu_j + \boldsymbol{\gamma}_j^T \boldsymbol{\nu}_y, \quad j = 1, \ldots, p,$$

where $\mu_j$ is the $j$th coordinate of $\boldsymbol{\mu}$ and $\boldsymbol{\gamma}_j^T$ is the $j$th row of $\boldsymbol{\Gamma}$. Under this setup $\boldsymbol{\Gamma}^T\mathbf{X}$ is again a sufficient reduction:

PROPOSITION 2. *Under the inverse exponential model* (3), *the distribution of $Y|\mathbf{X}$ is the same as the distribution of $Y|\boldsymbol{\Gamma}^T\mathbf{X}$ for all values of $\mathbf{X}$.*

Given $y$, $\boldsymbol{\mu}$ and $\boldsymbol{\Gamma}$ the likelihood for $\boldsymbol{\nu}_y$ is constructed from the products of (3) for $j = 1, \ldots, p$. This corresponds to fitting a generalized linear model with offsets $\boldsymbol{\mu}_j$, "predictors" $\boldsymbol{\gamma}_j$ and regression coefficient $\boldsymbol{\nu}_y$. The remaining parameters $\boldsymbol{\mu}$ and $\mathcal{S}_{\boldsymbol{\Gamma}} \in \mathcal{G}_{p \times d}$ can then be estimated by combining these intermediate partially maximized likelihoods over $y$. The essential point here is that *generalized PC models* can be constructed from the normal PC model (2) in the way that generalized linear models are constructed from linear models. Marx and Smith (1990) developed a method of principal component estimation for generalized linear models. Their approach, while recognizing issues that come with generalized linear models, seems quite different from the approach suggested here.

For example, if the coordinates of $\mathbf{X}_y$ are independent Bernoulli random variables, then we can express the model in terms of a multivariate logit defined coordinate-wise as $\operatorname{multlogit}_y = \boldsymbol{\mu} + \boldsymbol{\Gamma}\boldsymbol{\nu}_y$, where the right-hand side is the same as the mean function for the PC model (2). Since the predictors are conditionally independent, we can write

$$\Pr(\mathbf{X} = \mathbf{x}|Y = y) = \prod_{j=1}^p p_j(y)^{x_j} q_j(y)^{1-x_j},$$

where $\mathbf{x} = (x_j)$, $p_j(y) = \Pr(X_j = x_j|y)$, $q_j(y) = 1 - p_j(y)$ and $\log(p_j/q_j) = \mu_j + \boldsymbol{\gamma}_j^T\boldsymbol{\nu}_y$. The log likelihood

is therefore
$$\sum_y \left\{ \sum_{j=1}^p q_j(y) + \mathbf{x}_y^T(\boldsymbol{\mu} + \boldsymbol{\Gamma}\boldsymbol{\nu}_y) \right\},$$

which is to be maximized over $\boldsymbol{\mu} \in \mathbb{R}^p$, $\boldsymbol{\nu}_y \in \mathbb{R}^d$ and $\mathcal{S}_{\boldsymbol{\Gamma}} \in \mathcal{G}_{p \times d}$. By analogy with the normal case, this leads to *principal components for binary variables*, but they are not computed from the eigenvectors of a covariance matrix and they seem distinct from the recent proposal by de Leeuw (2006) for marginal reduction of $\mathbf{X}$.

## 4. PRINCIPAL FITTED COMPONENTS

In the PC model (2) no direct use is made of the response, which plays the role of an implicit conditioning argument. In principle, once the response is known, we should be able to tailor our reduction to that response. One way to adapt the reduction for a specific response is by modeling $\boldsymbol{\nu}_y$. This can be facilitated by graphical analyses when the response is bivariate or univariate. Recalling that $X_j$ denotes the $j$th predictor in $\mathbf{X}$, we can gain information from the data on the mean function $\mathrm{E}(X_j|Y)$ by investigating the $p$ two- or three-dimensional inverse response plots of $X_j$ versus $Y$, $j = 1, \ldots, p$, as described by Cook and Weisberg (1994, Chapter 8) and Cook (1998, Chapter 10). While such a graphical investigation might not be practical if $p$ is large, it might be doable when $p$ is in the tens, and certainly if $p$ is less than, say, 25.

Assume then that $\boldsymbol{\nu}_y = \boldsymbol{\beta}\mathbf{f}_y$, where $\boldsymbol{\beta} \in \mathbb{R}^{d \times r}$, $d \leq r$, has rank $d$ and $\mathbf{f}_y \in \mathbb{R}^r$ is a known vector-valued function of the response with $\sum_y \mathbf{f}_y = 0$. For example, if it is decided that each inverse mean function $\mathrm{E}(X_j|Y = y)$ can be modeled adequately by a cubic polynomial in $y$, then $\mathbf{f}_y$ equals $(y, y^2, y^3)^T$ minus its sample average. When $Y$ is univariate and graphical guidance is not available, $\mathbf{f}_y$ could be constructed by first partitioning the range of $Y$ into $h = r + 1$ "slices" or bins $H_k$, and then setting the $k$th coordinate $f_{yk}$ of $\mathbf{f}_y$ to

(4) $\quad f_{yk} = J(y \in H_k) - n_k/n, \quad k = 1, \ldots, r,$

where $J$ is the indicator function and $n_k$ is the number of observations falling in $H_k$. The $j$th coordinate $\nu_{yj}$ of $\boldsymbol{\nu}_y$ is then modeled as a constant in each slice $H_k$,

$$\nu_{yj} = \sum_{k=1}^r \beta_{jk} f_{yk} = \sum_{k=1}^r \beta_{jk}(J(y \in H_k) - n_k/n),$$

where $\beta_{jk}$ is the $jk$th element of $\boldsymbol{\beta}$. Each coordinate of the vector $\boldsymbol{\nu}_y = \boldsymbol{\beta}\mathbf{f}_y$ is now a step function that is constant within slices. Many other possibilities for basis functions are available in the literature. For example, we might adapt a classical Fourier series form (see, e.g., Eubank, 1988, page 82) and set $\mathbf{f}_y = \mathbf{g}_y - \bar{\mathbf{g}}$, where

$$\mathbf{g}_y = (\cos(2\pi y), \sin(2\pi y), \ldots,$$
$$\cos(2\pi k y), \sin(2\pi k y))^T$$

with $r = 2k$. We will use the slice basis function (4) later in this article because it leads to a connection with sliced inverse regression (SIR, Li, 1991). No claim is made that this is a generally reasonable nonparametric choice for $\mathbf{f}_y$. However, the slice basis function can be used to allow for replication in model (2), with the slices corresponding to the unique values of $y$. The parameterization $\boldsymbol{\nu}_y = \boldsymbol{\beta}\mathbf{f}_y$ can be used with exponential families as described in Section 3.3 and the PC model (2).

Substituting $\boldsymbol{\nu}_y = \boldsymbol{\beta}\mathbf{f}_y$ into model (2) we obtain the new model

(5) $\quad \mathbf{X}_y = \boldsymbol{\mu} + \boldsymbol{\Gamma}\boldsymbol{\beta}\mathbf{f}_y + \sigma\boldsymbol{\varepsilon},$

for which $\boldsymbol{\Gamma}^T\mathbf{X}$ is still a sufficient reduction. Let $\mathbb{X}$ denote the $n \times p$ matrix with rows $(\mathbf{X}_y - \bar{\mathbf{X}})^T$, let $\mathbf{F}$ denote the $n \times r$ matrix with rows $\mathbf{f}_y^T$, and let $\widehat{\mathbb{X}} = P_{\mathbf{F}}\mathbb{X}$ denote the $n \times p$ matrix of centered fitted values from the multivariate linear regression of $\mathbf{X}$ on $\mathbf{f}_y$, including an intercept. Holding $\boldsymbol{\Gamma}$ and $\sigma$ fixed at $\mathbf{G}$ and $s$, and maximizing the likelihood over $\boldsymbol{\mu}$ and $\boldsymbol{\beta}$, we obtain $\hat{\boldsymbol{\mu}} = \bar{\mathbf{X}}$, $\hat{\boldsymbol{\beta}} = \mathbf{G}^T\mathbb{X}^T\mathbf{F}(\mathbf{F}^T\mathbf{F})^{-1}$, and the partially maximized log likelihood (see Appendix A.3 for details)

(6) $\quad \begin{aligned} M_{\mathrm{PFC}}(\mathbf{G}, s^2) &= (-np/2)\log(s^2) \\ &\quad - (n/2s^2)\{\mathrm{trace}[\widehat{\boldsymbol{\Sigma}}] - \mathrm{trace}[P_{\mathbf{G}}\widehat{\boldsymbol{\Sigma}}_{\mathrm{fit}}]\}, \end{aligned}$

where $\widehat{\boldsymbol{\Sigma}}_{\mathrm{fit}} = \widehat{\mathbb{X}}^T\widehat{\mathbb{X}}/n$ is the sample covariance matrix of the fitted values. The likelihood again depends only on $\mathcal{S}_{\mathbf{G}}$. The rank of $\widehat{\boldsymbol{\Sigma}}_{\mathrm{fit}}$ is at most $r$ and typically $\mathrm{rank}(\widehat{\boldsymbol{\Sigma}}_{\mathrm{fit}}) = r$. In any event, we assume that $\mathrm{rank}(\widehat{\boldsymbol{\Sigma}}_{\mathrm{fit}}) \geq d$. The likelihood is then maximized by setting $\widehat{\mathcal{S}}_{\boldsymbol{\Gamma}}$ equal to the span of eigenvectors $\hat{\boldsymbol{\phi}}_1, \ldots, \hat{\boldsymbol{\phi}}_d$ corresponding to the largest $d$ eigenvalues $\hat{\lambda}_i^{\mathrm{fit}}, i = 1, \ldots, d$, of $\widehat{\boldsymbol{\Sigma}}_{\mathrm{fit}}$. The presence of replication in the $y$'s does not affect this estimator.

We call $\hat{\boldsymbol{\phi}}_1^T\mathbf{X}, \ldots, \hat{\boldsymbol{\phi}}_p^T\mathbf{X}$ *principal fitted components* (PFC), and call the associated eigenvectors $\hat{\boldsymbol{\phi}}_1, \ldots, \hat{\boldsymbol{\phi}}_d$

*PFC directions.* The corresponding estimate of scale is $\hat{\sigma}^2 = (\sum_{i=1}^{p} \hat{\lambda}_i - \sum_{i=1}^{d} \hat{\lambda}_i^{\text{fit}})/p$. A sufficient reduction under model (5) is then estimated by the first $d$ PFC's. We call model (5) a *PFC model* to distinguish it from the PC model (2).

## 5. COMPARING PC'S AND PFC'S

In the PC model (2) there are $(n-1)d$ $\nu$-parameters to be estimated, while in the PFC model (5) the corresponding number is just $dr$, which does not increase with $n$. Consequently, assuming that (5) is reasonable, we might expect it to yield more accurate estimates than (2).

### 5.1 Simulating $\widehat{\mathcal{S}}_{\Gamma}$

A small simulation using multivariate normal $(Y, \mathbf{X})$ was conducted to obtain first insights into the operating characteristics of principal components and principal fitted components. Here and in all other simulations we restrict $\boldsymbol{\Gamma} \in \mathbb{R}^p$ ($d = 1$) because this allows straightforward comparisons with forward OLS. The two component methods are applicable when $d > 1$, but then the OLS fit of $Y$ on $\mathbf{X}$ must necessarily miss $d-1$ directions. First, $Y$ was generated as a normal random variable with mean 0 and variance $\sigma_Y^2$, and then $\mathbf{X}_y$ was generated according to the inverse model

$$(7) \qquad \mathbf{X}_y = \boldsymbol{\Gamma} y + \sigma \boldsymbol{\varepsilon},$$

with $\boldsymbol{\Gamma} = (1, 0, \ldots, 0)^T$, $p = 10$ and $\sigma > 0$. The forward regression $Y|\mathbf{X}$ follows a textbook normal linear regression model,

$$(8) \qquad Y = \alpha_0 + \boldsymbol{\alpha}^T \mathbf{x} + \sigma_{Y|\mathbf{X}} \epsilon,$$

where $\mathbf{x}$ denotes an observed value of $\mathbf{X}$, $\sigma_{Y|\mathbf{X}}$ is constant, $\epsilon$ is a standard normal random variable and, as indicated by Proposition 1 and its extension to PFC models, $\text{span}(\boldsymbol{\alpha}) = \text{span}(\boldsymbol{\Gamma})$. Thus, $E(Y|\mathbf{X})$ depends on $\mathbf{X}$ via $\boldsymbol{\Gamma}^T \mathbf{X}$, which is equal to $X_1$ for this simulation. Let $\widehat{\boldsymbol{\alpha}}$ denote the OLS estimator of $\boldsymbol{\alpha}$. We consider three ways of estimating $\mathcal{S}_{\Gamma}$, each based on a correct model: OLS, using $\text{span}(\widehat{\boldsymbol{\alpha}})$; PC, using $\text{span}(\hat{\boldsymbol{\gamma}}_1)$; and PFC, using $\text{span}(\hat{\boldsymbol{\phi}}_1)$ from the fit of model (5) with $\mathbf{f}_y = y - \bar{y}$.

Each data set was summarized conveniently by computing the angle between $\boldsymbol{\Gamma}$ and each of $\hat{\boldsymbol{\gamma}}_1$ (PC), $\hat{\boldsymbol{\phi}}_1$ (PFC) and $\widehat{\boldsymbol{\alpha}}$ (OLS). Other summary measures were tested but they left the same qualitative impressions as the angle. The mean squared error (9) used in Figure 1(d) and discussed in Section 5.2 is one instance of an alternative summary measure. Three aspects of the simulation model were varied, the sample size $n$, the conditional error standard deviation $\sigma$ and the marginal standard deviation of $Y$, $\sigma_Y$. It might be expected that PFC will do better than PC in this simulation. While both methods are based on a correct model, the PFC model is more parsimonious. On the other hand, it could be more difficult to anticipate a relation between these inverse methods and OLS, which is based on a correct forward model.

Shown in Figure 1(a) are average angles taken over 500 replications versus $n$, with $\sigma = \sigma_Y = 1$. On the average, the OLS vector was observed to be a bit closer to $\mathcal{S}_{\Gamma}$ than the PC vector $\hat{\boldsymbol{\gamma}}_1$, except for small samples, when $\hat{\boldsymbol{\gamma}}_1$ was the better of the two. More importantly, the PFC vector $\hat{\boldsymbol{\phi}}_1$ was observed to be more accurate than the other two estimators at all sample sizes. The difference between the PFC and OLS estimators can be increased by varying $\sigma_Y$, as is apparent from Figure 1(b).

Figure 1(b) shows the results of a second series of simulations at various values of $\sigma_Y$ with $n = 40$ and $\sigma = 1$. The method of principal fitted components is seen to be superior for small values of $\sigma_Y$, while it is essentially equivalent to principal components for large values. Perhaps surprisingly, the OLS estimator is clearly the worst method over most of the range of $\sigma_Y$. Figure 1(c) shows average angles as $\sigma$ varies with $n = 40$ and $\sigma_Y = 1$. Again, PFC is seen to be the best method, with the relative performance of PC and PFC depending on the value of $\sigma$. The accuracy of the PC or PFC estimates of $\mathcal{S}_{\Gamma}$ should improve as $n$ increases, or as $\sigma_Y$ increases or as $\sigma$ decreases. The results in Figure 1 agree with this expectation.

To explain the relative behavior of the OLS estimates in the simulation results for model (7), let $R = \sigma_Y^2 / (\sigma_Y^2 + \sigma^2)$. Then it can be shown that $\boldsymbol{\alpha} = R\boldsymbol{\Gamma}$, that $\sqrt{n}(\widehat{\boldsymbol{\alpha}} - \boldsymbol{\alpha})$ is asymptotically normal with mean 0 and covariance matrix

$$\text{Var}(\widehat{\boldsymbol{\alpha}}) = RQ_{\boldsymbol{\Gamma}} + R(1-R)P_{\boldsymbol{\Gamma}},$$

and that $\boldsymbol{\alpha}^T [\text{Var}(\widehat{\boldsymbol{\alpha}})]^{-1} \boldsymbol{\alpha} = R/(1-R)$. As $\sigma_Y \to \infty$, $R \to 1$ but $\text{Var}(\widehat{\boldsymbol{\alpha}}) \geq RQ_{\boldsymbol{\Gamma}}$. Thus, $\text{Var}(\widehat{\boldsymbol{\alpha}})$ is bounded below, and this explains the behavior of the OLS estimates in Figure 1(b). On the other hand, as $\sigma \to \infty$, $\boldsymbol{\alpha} \to 0$ and $\text{Var}(\widehat{\boldsymbol{\alpha}}) \to 0$, but $\boldsymbol{\alpha}^T [\text{Var}(\widehat{\boldsymbol{\alpha}})]^{-1} \boldsymbol{\alpha} \to 0$ as well. Consequently, $\boldsymbol{\alpha} \to 0$ faster than the "standard deviation" of its estimates and the performance of $\widehat{\boldsymbol{\alpha}}$ must deteriorate. These results can be extended straightforwardly to model (2) with $d = 1$, but then the behavior of the OLS estimator will depend on $\text{Var}(\boldsymbol{\nu}_Y)$ and $\text{Cov}(\boldsymbol{\nu}_Y, Y)$.

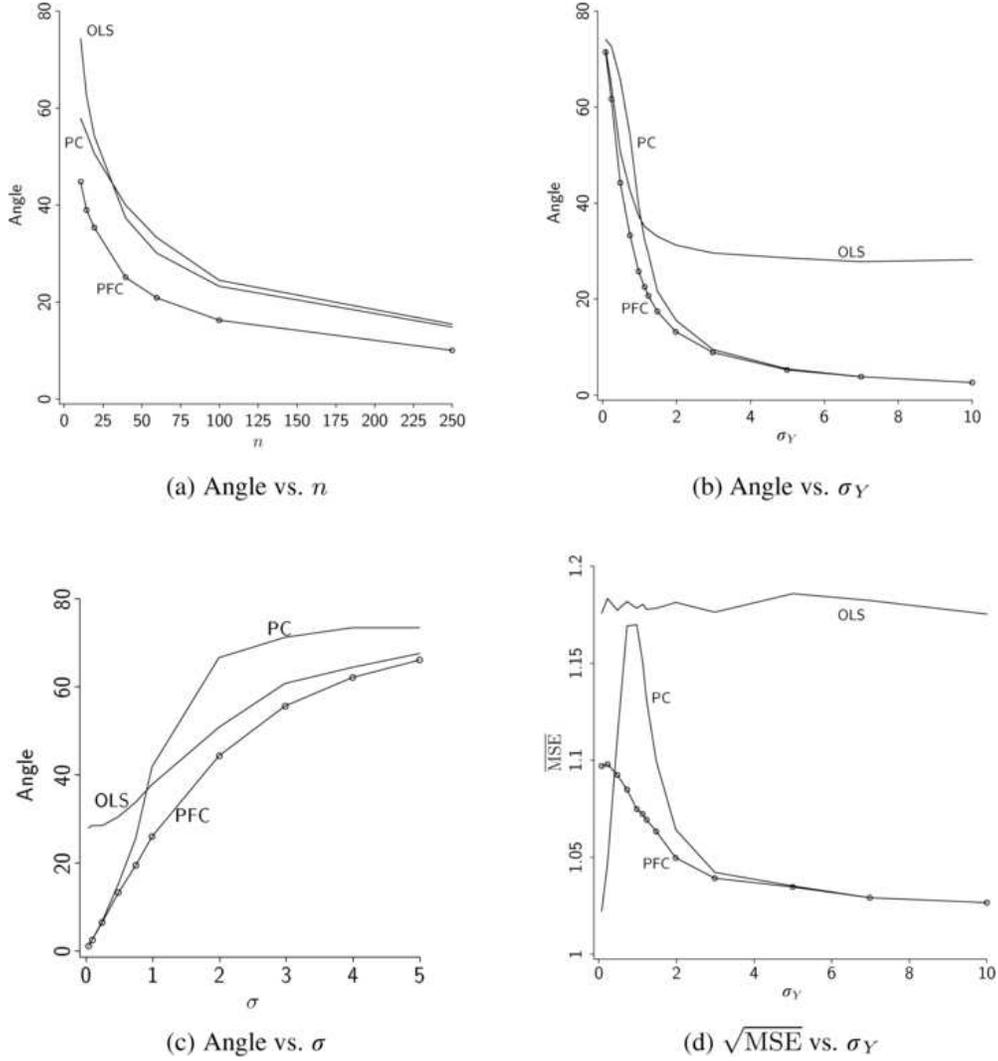

FIG. 1. *Simulation results for model* (7). *(a)–(c) display average simulation angles between the estimated and the true direction versus* (a) *sample size with* $\sigma_Y = \sigma = 1$; (b) $\sigma_Y$ *with* $n = 40$, $\sigma = 1$; *and* (c) $\sigma$ *with* $n = 40$, $\sigma_Y = 1$. (d) *is the square root of the standardized MSE of prediction versus* $\sigma_Y$ *with* $n = 40$ *and* $\sigma = 1$.

Letting $\Gamma_k$ denote the $k$th element of the $\boldsymbol{\Gamma}$ in model (7), the marginal correlation between the $j$th and $k$th predictors, $j \neq k$, is

$$\rho_{jk} = \frac{\Gamma_k \Gamma_j \sigma_Y^2}{\sqrt{(\sigma^2 + \Gamma_j^2 \sigma_Y^2)(\sigma^2 + \Gamma_k^2 \sigma_Y^2)}}.$$

If $\Gamma_j$ and $\Gamma_k$ are both nonzero, then $|\rho_{jk}| \to 1$ as $\sigma_Y^2 \to \infty$ with $\sigma^2$ fixed. However, $\boldsymbol{\Gamma} = (1, 0, \ldots, 0)^T$ in the version of model (7) used to produce Figure 1(b), with the consequence that the predictors are both marginally and conditionally independent. Thus, the bottoming out of the OLS estimator in Figure 1(b) for large values of $\sigma_Y$ and in Figure 1(c) for small values of $\sigma$ has nothing to do with collinearity. To gain further insights into the impact of collinearity in this situation, we repeated the simulation leading to Figure 1(b) with $\boldsymbol{\Gamma} = (1, \ldots, 1)^T / \sqrt{10}$. Now, $\rho_{jk} \to 1$ as $\sigma_Y \to \infty$ for all pairs of predictors. However, within the simulation error, these new results were identical to those of Figure 1(b). This suggests that the value of principal component estimators does not rest solely with the presence of collinearity.

## 5.2 Prediction

While this article is focused on estimation of reductive subspaces, this first simulation study provides a convenient place to touch base with predictive considerations. Since the forward regression follows a normal linear model, we characterize predictive performance by using the scaled mean squared error

$$(9) \quad \text{MSE} = \text{E}(Y_f - \hat{a} - \hat{b}\widehat{\boldsymbol{\Gamma}}^T\mathbf{X}_f)^2/\sigma^2_{Y|\mathbf{X}},$$

where $(Y_f, \mathbf{X}_f)$ represents a future observation on $(Y, \mathbf{X})$, and the expectation is taken over $(Y_f, \mathbf{X}_f)$ and the data. The forward model error variance $\sigma^2_{Y|\mathbf{X}}$ was used for scaling because the numerator of (9) is in the units of $Y^2$, and the MSE will be constant for OLS as we vary $\sigma_Y$. $\widehat{\boldsymbol{\Gamma}}$ can be $\hat{\boldsymbol{\gamma}}_1$ (PC), $\hat{\boldsymbol{\phi}}_1$ (PFC) or $\widehat{\boldsymbol{\alpha}}$ (OLS). The intercept $\hat{a}$ and slope $\hat{b}$ were then computed from the OLS fit of $Y_i$ on $\widehat{\boldsymbol{\Gamma}}^T\mathbf{X}_i$ ($\hat{b} = 1$ for $\widehat{\boldsymbol{\alpha}}$). The MSE, which is bounded below by 1, was estimated by first calculating the expectation explicitly over the future observations and then using 500 simulated data sets to estimate the remaining expectation over the data. Figure 1(d) shows the resulting MSE as a function of $\sigma_Y$ and the three estimators. The PFC estimator performed the best, except for small values of $\sigma_Y$, where the PC estimator did better. The OLS estimator was dominated by the other two.

## 6. INVERSE MODELS WITH STRUCTURED ERRORS

Let $\boldsymbol{\Gamma}_0 \in \mathbb{R}^{p \times (p-d)}$ denote a *completion* of $\boldsymbol{\Gamma}$; that is, $(\boldsymbol{\Gamma}_0, \boldsymbol{\Gamma}) \in \mathbb{R}^{p \times p}$ is an orthogonal matrix. The PC model (2) and the PFC model (5) have the property that $Y$ is independent of $\boldsymbol{\Gamma}_0^T\mathbf{X}$ both marginally, $Y \perp\!\!\!\perp \boldsymbol{\Gamma}_0^T\mathbf{X}$, and conditionally, $Y \perp\!\!\!\perp \boldsymbol{\Gamma}_0^T\mathbf{X}|\boldsymbol{\Gamma}^T\mathbf{X}$. Consequently, $(Y, \boldsymbol{\Gamma}^T\mathbf{X}) \perp\!\!\!\perp \boldsymbol{\Gamma}_0^T\mathbf{X}$. This enables us to identify $\boldsymbol{\Gamma}_0^T\mathbf{X}$ unambiguously as *irrelevant information* to be excluded (Fisher, 1922). The PC and PFC models also have quite restrictive conditions on $\text{Var}(\mathbf{X}_y) = \sigma^2 I_p$. In this section we extend the variance function of these models while preserving the form of the relevant and irrelevant information.

### 6.1 An Extended PC Model

Consider the PC model with heterogeneous errors,

$$(10) \quad \mathbf{X}_y = \boldsymbol{\mu} + \boldsymbol{\Gamma}\boldsymbol{\nu}_y + \boldsymbol{\Gamma}_0\boldsymbol{\Omega}_0\boldsymbol{\varepsilon}_0 + \boldsymbol{\Gamma}\boldsymbol{\Omega}\boldsymbol{\varepsilon},$$

where $\boldsymbol{\mu}$, $\boldsymbol{\Gamma}$, $\boldsymbol{\Gamma}_0$ and $\boldsymbol{\nu}_y$ are as defined previously. The error vectors $\boldsymbol{\varepsilon}_0 \in \mathbb{R}^{p-d}$ and $\boldsymbol{\varepsilon} \in \mathbb{R}^d$ are independent and normally distributed, each with mean 0 and identity covariance matrix. The full-rank matrices $\boldsymbol{\Omega} \in \mathbb{R}^{d \times d}$ and $\boldsymbol{\Omega}_0 \in \mathbb{R}^{(p-d) \times (p-d)}$ serve in part to convert the normal errors to appropriate scales and, without loss of generality, are assumed to be symmetric. Since the two error components must be independent, the variance structure in this model is still restrictive, but it is considerably more permissive than the variance structure in the first PC model (2) and the components of $\mathbf{X}$ can now be in different scales.

Linearly transforming both sides of model (10), we have $\boldsymbol{\Gamma}^T\mathbf{X}_y = \boldsymbol{\Gamma}^T\boldsymbol{\mu} + \boldsymbol{\nu}_y + \boldsymbol{\Omega}\boldsymbol{\varepsilon}$ and $\boldsymbol{\Gamma}_0^T\mathbf{X}_y = \boldsymbol{\Gamma}_0^T\boldsymbol{\mu} + \boldsymbol{\Omega}_0\boldsymbol{\varepsilon}_0$. From these representations we see that $(\boldsymbol{\Gamma}, \boldsymbol{\Gamma}_0)^T\mathbf{X}_y$ contains two independent components. The active $y$-dependent component $\boldsymbol{\Gamma}^T\mathbf{X}_y$ consists of the coordinates of the projection of $\mathbf{X}$ onto $\mathcal{S}_{\boldsymbol{\Gamma}}$. It has an arbitrary mean and constant but a general covariance matrix. The other projective component lives in the orthogonal complement $\text{span}(\boldsymbol{\Gamma}_0) = \mathcal{S}_{\boldsymbol{\Gamma}}^\perp$ and has constant mean and variance. Most importantly, this extended model preserves the independence property that $(Y, \boldsymbol{\Gamma}^T\mathbf{X}) \perp\!\!\!\perp \boldsymbol{\Gamma}_0^T\mathbf{X}$, so $\boldsymbol{\Gamma}^T\mathbf{X}$ is a sufficient reduction:

PROPOSITION 3. *Under the extended PC model* (10), *the distribution of* $Y|\mathbf{X}$ *is the same as the distribution of* $Y|\boldsymbol{\Gamma}^T\mathbf{X}$ *for all values of* $\mathbf{X}$.

Turning to estimation, we assume that the $y$'s are distinct. Replication for model (10) can be addressed with a version of model (13), which is discussed in Section 6.2, by using the slice basis function for $\mathbf{f}_y$ to indicate identical $y$'s. Maximizing over $\boldsymbol{\mu}$, $\boldsymbol{\nu}_y$ and $\boldsymbol{\Omega}_0$, the partially maximized log likelihood $L_{\text{PC}}$ is a function of possible values $\mathbf{G}$ and $\mathbf{M}$ for $\boldsymbol{\Gamma}$ and $\boldsymbol{\Omega}^2$, apart from constants:

$$L_{\text{PC}}(\mathbf{G}, \mathbf{M})$$
$$= -(n/2)\log(|\mathbf{G}_0^T\widehat{\boldsymbol{\Sigma}}\mathbf{G}_0|) - (n/2)\log(|\mathbf{M}|),$$

where $\mathbf{G}_0$ is a completion of $\mathbf{G}$. $\boldsymbol{\Omega}^2$ is not estimable, but because the likelihood factors, it is maximized over $\mathbf{G}_0$ for any fixed $\mathbf{M}$ by any full-rank linear transformation of the last $p - d$ sample PC directions. Consequently, we might be tempted to take

$$(11) \quad \widehat{\mathcal{S}}_{\boldsymbol{\Gamma}} = \text{span}^\perp(\hat{\boldsymbol{\gamma}}_{d+1}, \ldots, \hat{\boldsymbol{\gamma}}_p).$$

Thus a sufficient reduction is again estimated by the first $d$ sample principal components computed from $\widehat{\boldsymbol{\Sigma}}$. Nevertheless, additional conditions are necessary for principal components to work well under this model.

Some of the differences between the PC model (2) and the extended version (10) can be seen in the marginal covariances of the predictors. Under model (2)

$$\boldsymbol{\Sigma} = \sigma^2 \boldsymbol{\Gamma}_0 \boldsymbol{\Gamma}_0^T + \boldsymbol{\Gamma}\{\sigma^2 I_d + \text{Var}(\boldsymbol{\nu}_Y)\}\boldsymbol{\Gamma}^T.$$

Here the smallest eigenvalue of $\boldsymbol{\Sigma}$ is equal to $\sigma^2$ with multiplicity $p - d$ and corresponding eigenvectors $\boldsymbol{\Gamma}_0$. The largest $d$ eigenvalues, which are all larger than $\sigma^2$, are the same as the eigenvalues of $\sigma^2 I_d + \text{Var}(\boldsymbol{\nu}_Y)$. The corresponding PC directions are $\boldsymbol{\Gamma}\mathbf{v}_j$, $j = 1, \ldots, d$, where $\{\mathbf{v}_j\}$ are the eigenvectors of $\sigma^2 I_d + \text{Var}(\boldsymbol{\nu}_Y)$. In this case $\mathcal{S}_{\boldsymbol{\Gamma}}$ and $\mathcal{S}_{\boldsymbol{\Gamma}_0}$ are distinguished by the magnitudes of the corresponding eigenvalues of $\boldsymbol{\Sigma}$, providing the likelihood information to identify $\mathcal{S}_{\boldsymbol{\Gamma}}$.

Under model (10)

$$\boldsymbol{\Sigma} = \boldsymbol{\Gamma}_0 \boldsymbol{\Omega}_0^2 \boldsymbol{\Gamma}_0^T + \boldsymbol{\Gamma}\{\boldsymbol{\Omega}^2 + \text{Var}(\boldsymbol{\nu}_Y)\}\boldsymbol{\Gamma}^T$$
$$= \boldsymbol{\Gamma}_0 \mathbf{V}_0 \mathbf{D}_0 \mathbf{V}_0^T \boldsymbol{\Gamma}_0^T + \boldsymbol{\Gamma}\mathbf{V}\mathbf{D}\mathbf{V}^T \boldsymbol{\Gamma}^T,$$

where $\mathbf{V}_0 \mathbf{D}_0 \mathbf{V}_0^T$ and $\mathbf{V}\mathbf{D}\mathbf{V}^T$ are the spectral decompositions of $\boldsymbol{\Omega}_0^2$ and $\boldsymbol{\Omega}^2 + \text{Var}(\boldsymbol{\nu}_Y)$. The PC directions under model (10) can be written *unordered* as $\boldsymbol{\Gamma}_0 \mathbf{V}_0$ and $\boldsymbol{\Gamma}\mathbf{V}$ with eigenvalues given by the corresponding elements of the diagonal matrices $\mathbf{D}_0$ and $\mathbf{D}$. The estimate of $\mathcal{S}_{\boldsymbol{\Gamma}}$ given in (11) will be consistent if the largest eigenvalue in $\mathbf{D}_0$ is smaller than the smallest eigenvalue in $\mathbf{D}$. In other words, the likelihood-based estimator (11) should be reasonable if the signal as represented by $\mathbf{D}$ is larger than the noise as represented by $\mathbf{D}_0$.

To help fix ideas, consider the extension of model (7),

(12) $\quad \mathbf{X}_y = \boldsymbol{\Gamma} y + \sigma_0 \boldsymbol{\Gamma}_0 \boldsymbol{\varepsilon}_0 + \sigma \boldsymbol{\Gamma}\boldsymbol{\varepsilon},$

where $(Y, \mathbf{X})$ is normally distributed, and the sufficient reduction still has dimension 1. The forward regression $Y|\mathbf{X}$ follows a normal linear regression model where the mean function depends on $\mathbf{X}$ only via $\boldsymbol{\Gamma}^T \mathbf{X}$, and

$$\boldsymbol{\Sigma} = \sigma_0^2 \boldsymbol{\Gamma}_0 \boldsymbol{\Gamma}_0^T + (\sigma_Y^2 + \sigma^2)\boldsymbol{\Gamma}\boldsymbol{\Gamma}^T.$$

If $\sigma_0^2 < \sigma_Y^2 + \sigma^2$, then the first population principal component yields a sufficient reduction $\boldsymbol{\Gamma}^T \mathbf{X}$. If $\sigma_0^2 > \sigma_Y^2 + \sigma^2$, then a sufficient reduction is given by the last population principal component, illustrating the possibility suggested by Cox (1968). However, if $\sigma_0^2 = \sigma_Y^2 + \sigma^2$, then principal components will fail since all of the eigenvalues of $\boldsymbol{\Sigma}$ are equal.

In short, principal components may yield reasonable reductions if the signal dominates the noise in the extended PC model (10).

### 6.2 An Extended PFC Model

In this section we consider a PFC model with heterogeneous errors by modeling the $\nu$-parameters in (10),

(13) $\quad \mathbf{X}_y = \boldsymbol{\mu} + \boldsymbol{\Gamma}\boldsymbol{\beta}\mathbf{f}_y + \boldsymbol{\Gamma}_0 \boldsymbol{\Omega}_0 \boldsymbol{\varepsilon}_0 + \boldsymbol{\Gamma}\boldsymbol{\Omega}\boldsymbol{\varepsilon},$

where all terms are as defined previously, except we no longer require that $d \leq r$. The maximum likelihood estimators under (13) are not affected by the presence of replication in the $y$'s. Extensions of the exponential family model (3) to permit dependence among the predictors may depend on the particular family involved. For instance, the quadratic exponential model described by Zhao and Prentice (1990) might be useful when the predictors are correlated binary variables.

Under model (13), $\boldsymbol{\Gamma}^T \mathbf{X}$ is again a sufficient reduction, so we are still interested in estimating the reductive subspace $\mathcal{S}_{\boldsymbol{\Gamma}}$. Maximizing the log likelihood over $\boldsymbol{\mu}$ and $\boldsymbol{\beta}$, we find the partially maximized log likelihood, apart from unimportant constants,

$$-(n/2)\log|\mathbf{M}_0| - (n/2)\,\text{trace}[\mathbf{M}_0^{-1}\mathbf{G}_0^T \mathbb{X}^T \mathbb{X}\mathbf{G}_0/n]$$
$$- (n/2)\log|\mathbf{M}|,$$
$$-(n/2)\,\text{trace}[\mathbf{M}^{-1}\mathbf{G}^T\{\mathbb{X}^T \mathbb{X} - \mathbb{X}^T P_{\mathbf{F}} \mathbb{X}\}\mathbf{G}/n],$$

where $\mathbf{G}$, $\mathbf{M}$ and $\mathbf{M}_0$ represent possible values for $\boldsymbol{\Gamma}$, $\boldsymbol{\Omega}^2$ and $\boldsymbol{\Omega}_0^2$. After maximizing over $\mathbf{M}$ and $\mathbf{M}_0$ (see, e.g., Muirhead, 1982, page 84), the maximum likelihood estimate of $\mathcal{S}_{\boldsymbol{\Gamma}}$ is found by maximizing

(14)
$$L_{\text{PFC}}(\mathbf{G}) = (-n/2)\log|\mathbf{G}_0^T \widehat{\boldsymbol{\Sigma}}\mathbf{G}_0|$$
$$- (n/2)\log|\mathbf{G}^T \widehat{\boldsymbol{\Sigma}}_{\text{res}}\mathbf{G}|,$$

where $\widehat{\boldsymbol{\Sigma}}_{\text{res}} = \widehat{\boldsymbol{\Sigma}} - \widehat{\boldsymbol{\Sigma}}_{\text{fit}}$ is the sample covariance matrix of the residuals from the fit of $\mathbf{X}_y$ on $\mathbf{f}_y$. Like previous likelihoods, (14) depends only on $\mathcal{S}_{\mathbf{G}}$. But in contrast to the previous likelihoods, here there does not seem to be a recognizable estimate for $\mathcal{S}_{\boldsymbol{\Gamma}}$, and thus $\widehat{\mathcal{S}}_{\boldsymbol{\Gamma}}$ must be obtained by maximizing (14) numerically over the Grassmann manifold $\mathcal{G}_{p \times d}$.

The following proposition gives population versions of matrices in the partially maximized likelihood function (14). It summarizes some of the discussion in Section 6.1, and provides results that will be useful for studying (14).

PROPOSITION 4. *Assume the extended PFC model* (13) *with uncorrelated but not necessarily normal errors,* $\mathrm{Var}((\boldsymbol{\varepsilon}_0^T, \boldsymbol{\varepsilon}^T)^T) = I_p$. *Then*

$$\widehat{\boldsymbol{\Sigma}} \xrightarrow{p} \boldsymbol{\Sigma} = \boldsymbol{\Gamma}_0 \boldsymbol{\Omega}_0^2 \boldsymbol{\Gamma}_0^T + \boldsymbol{\Gamma}\{\boldsymbol{\Omega}^2 + \boldsymbol{\beta}\mathrm{Var}(\mathbf{f}_Y)\boldsymbol{\beta}^T\}\boldsymbol{\Gamma}^T,$$

$$\widehat{\boldsymbol{\Sigma}}_{\mathrm{fit}} \xrightarrow{p} \boldsymbol{\Sigma}_{\mathrm{fit}} = \boldsymbol{\Gamma}\boldsymbol{\beta}\mathrm{Var}(\mathbf{f}_Y)\boldsymbol{\beta}^T\boldsymbol{\Gamma}^T,$$

$$\widehat{\boldsymbol{\Sigma}}_{\mathrm{res}} \xrightarrow{p} \boldsymbol{\Sigma}_{\mathrm{res}} = \boldsymbol{\Gamma}_0\boldsymbol{\Omega}_0^2\boldsymbol{\Gamma}_0^T + \boldsymbol{\Gamma}\boldsymbol{\Omega}^2\boldsymbol{\Gamma}^T.$$

To understand the behavior of the function $L_{\mathrm{PFC}}(\mathbf{G})$ (14) in a bit more detail, write it in the form, with $\mathbf{H} = P_{\mathbf{F}}\mathbb{X}\mathbf{G}/\sqrt{n}$,

$$-2L_{\mathrm{PFC}}(\mathbf{G})/n$$
$$= \log|\mathbf{G}_0^T\widehat{\boldsymbol{\Sigma}}\mathbf{G}_0| + \log|\mathbf{G}^T\widehat{\boldsymbol{\Sigma}}\mathbf{G} - \mathbf{G}^T\mathbb{X}^TP_{\mathbf{F}}\mathbb{X}\mathbf{G}/n|$$
$$= \log\{|\mathbf{G}_0^T\widehat{\boldsymbol{\Sigma}}\mathbf{G}_0||\mathbf{G}^T\widehat{\boldsymbol{\Sigma}}\mathbf{G}||I_n - \mathbf{H}(\mathbf{G}^T\widehat{\boldsymbol{\Sigma}}\mathbf{G})^{-1}\mathbf{H}^T|\}$$
$$= \log\{|\mathbf{G}_0^T\widehat{\boldsymbol{\Sigma}}\mathbf{G}_0||\mathbf{G}^T\widehat{\boldsymbol{\Sigma}}\mathbf{G}||I_n - P_{\mathbf{F}}P_{\mathbb{X}\mathbf{G}}P_{\mathbf{F}}|\}.$$

The first product in the log is such that

$$(15) \qquad |\mathbf{G}_0^T\widehat{\boldsymbol{\Sigma}}\mathbf{G}_0||\mathbf{G}^T\widehat{\boldsymbol{\Sigma}}\mathbf{G}| \geq |\widehat{\boldsymbol{\Sigma}}|,$$

and it achieves its lower bound when the columns of $\mathbf{G}$ are *any* $d$ sample PC directions and the columns of $\mathbf{G}_0$ consist of the remaining directions. Consequently, the function $-\log\{|\mathbf{G}_0^T\widehat{\boldsymbol{\Sigma}}\mathbf{G}_0||\mathbf{G}^T\widehat{\boldsymbol{\Sigma}}\mathbf{G}|\}$ has at least $\binom{p}{d}$ local maxima of equal height. It is then up to the last term $-\log|I_n - P_{\mathbf{F}}P_{\mathbb{X}\mathbf{G}}P_{\mathbf{F}}|$ to reshape $L_{\mathrm{PFC}}$ into a possibly multimodal surface over $\mathcal{G}_{p \times d}$ with a single global maximum. If a $\mathbf{G}$ can be found so that $\mathrm{span}(\mathbb{X}\mathbf{G}) \subseteq \mathrm{span}(\mathbf{F})$, then $|I_n - P_{\mathbf{F}}P_{\mathbb{X}\mathbf{G}}P_{\mathbf{F}}| = 0$ and the log likelihood is infinite at its maximum. If the signal is weak, or in the extreme $\mathbf{F}^T\mathbb{X} \approx 0$, then $|I_n - P_{\mathbf{F}}P_{\mathbb{X}\mathbf{G}}P_{\mathbf{F}}| \approx 1$ for all $\mathbf{G}$, this term will contribute little to the log likelihood, and we will be left with a surface having perhaps many local maxima of similar heights.

Because the likelihood surface may be multimodal, use of standard gradient optimization methods may be problematic in some regressions. Consideration of a candidate set of solutions or starting values might mitigate the problems resulting from a multimodal surface and can further illuminate the role of principal components. It also provides a relatively straightforward way to explore this methodology if computer code for Grassmann optimization is not conveniently available.

Candidate sets can be constructed using the following rationale. Recall from the discussion in Section 6.1 that the unordered PC directions are of the form $(\boldsymbol{\Gamma}_0\mathbf{V}_0, \boldsymbol{\Gamma}\mathbf{V})$, and they comprise one possible population candidate set. This structure means that there is a subset of $d$ PC directions $\boldsymbol{\gamma}_{(1)}, \ldots, \boldsymbol{\gamma}_{(d)}$, such that

$$\mathcal{S}_{\boldsymbol{\Gamma}} = \mathrm{span}(\boldsymbol{\Gamma}\mathbf{V}) = \mathrm{span}\{\boldsymbol{\gamma}_{(1)}, \ldots, \boldsymbol{\gamma}_{(d)}\}.$$

Consequently, provided that the eigenvalues corresponding to $\boldsymbol{\Gamma}\mathbf{V}$ are distinct from those corresponding to $\boldsymbol{\Gamma}_0\mathbf{V}_0$, we can construct an estimator of the sufficient reduction $\mathbf{V}^T\boldsymbol{\Gamma}^T\mathbf{X}$ by evaluating $L_{\mathrm{PFC}}(\mathbf{G})$ at all subsets of $d$ sample PC directions $\hat{\boldsymbol{\gamma}}_{(1)}, \ldots, \hat{\boldsymbol{\gamma}}_{(d)}$, and then choosing the subset with the highest likelihood. We call this the $\mathrm{PFC}_{\mathrm{PC}}$ method, because the PFC likelihood is being evaluated at the PC directions. This method might be awkward to implement if the number of combinations to be evaluated is large. As an alternative, we could choose PC directions sequentially:

1. Find $\hat{\boldsymbol{\gamma}}_{(1)} = \arg\max L_{\mathrm{PFC}}(\mathbf{h})$, where the maximum is taken over the $p \times 1$ vector $\mathbf{h}$ in the PC candidate set $A = \{\hat{\boldsymbol{\gamma}}_j, j = 1, \ldots, p\}$.
2. Find $\hat{\boldsymbol{\gamma}}_{(2)} = \arg\max L_{\mathrm{PFC}}(\hat{\boldsymbol{\gamma}}_{(1)}, \mathbf{h})$, where the maximum is now taken over the $p \times 1$ vector $\mathbf{h}$ in the reduced candidate set $A - \{\hat{\boldsymbol{\gamma}}_{(1)}\}$.
3. Continue until reaching the final maximization

$$\hat{\boldsymbol{\gamma}}_{(d)} = \arg\max L_{\mathrm{PFC}}(\hat{\boldsymbol{\gamma}}_{(1)}, \ldots, \hat{\boldsymbol{\gamma}}_{(d-1)}, \mathbf{h}),$$

where the maximum is now taken over $\mathbf{h}$ in $A - \{\hat{\boldsymbol{\gamma}}_{(1)}, \ldots, \hat{\boldsymbol{\gamma}}_{(d-1)}\}$.

This approach is similar in spirit to other methodology for selecting principal components (Jolliffe, 2002), but here we base the selection on a likelihood.

Using Proposition 4 and following the same rationale we can construct two additional candidate sets. One consists of the PFC directions, the eigenvectors of $\widehat{\boldsymbol{\Sigma}}_{\mathrm{fit}}$, and the other contains the eigenvectors of $\widehat{\boldsymbol{\Sigma}}_{\mathrm{res}}$, which are called the *residual component* ($RC$) *directions*. The estimator constructed by evaluating the PFC likelihood (14) at all subsets of the candidate set consisting of the PC, PFC and RC directions will be denoted as $\mathrm{PFC}_{\mathrm{all}}$.

Before turning again to illustrative simulations, we connect principal fitted components with sliced inverse regression.

### 6.3 PFC and SIR

To relate SIR and PFC, $\mathbf{f}_y$ *must* be constructed using the slice basis function (4). Let $\bar{\mathbf{X}}_k = \sum_{i \in H_k} \mathbf{X}_i/n_k$ and $\bar{\mathbf{X}} = \sum_{i=1}^n \mathbf{X}_i/n$. Then it can be shown that $\widehat{\boldsymbol{\Sigma}}_{\mathrm{fit}}$ is the sample covariance matrix of the slice means

$\bar{\mathbf{X}}_k$,

$$\widehat{\boldsymbol{\Sigma}}_{\text{fit}} = \sum_{k=1}^{h} (n_k/n)(\bar{\mathbf{X}}_k - \bar{\mathbf{X}})(\bar{\mathbf{X}}_k - \bar{\mathbf{X}})^T$$
$$= \widehat{\boldsymbol{\Sigma}}^{1/2} \widehat{\boldsymbol{\Sigma}}_{\text{sir}} \widehat{\boldsymbol{\Sigma}}^{1/2},$$

where $\widehat{\boldsymbol{\Sigma}}_{\text{sir}} = \widehat{\boldsymbol{\Sigma}}^{-1/2} \widehat{\boldsymbol{\Sigma}}_{\text{fit}} \widehat{\boldsymbol{\Sigma}}^{-1/2}$ is the usual SIR kernel matrix in the standardized scale of $\mathbf{Z} = \boldsymbol{\Sigma}^{-1/2}(\mathbf{X} - \mathrm{E}(\mathbf{X}))$, with sample version $\hat{\mathbf{Z}} = \widehat{\boldsymbol{\Sigma}}^{-1/2}(\mathbf{X} - \bar{\mathbf{X}})$. Substituting this form into the partially maximized log likelihood (14), we have

$$-2L_{\text{PFC}}(\mathbf{G})/n = \log|\mathbf{G}_0^T \widehat{\boldsymbol{\Sigma}} \mathbf{G}_0|$$
$$+ \log|\mathbf{G}^T \widehat{\boldsymbol{\Sigma}}^{1/2} \{I_p - \widehat{\boldsymbol{\Sigma}}_{\text{sir}}\} \widehat{\boldsymbol{\Sigma}}^{1/2} \mathbf{G}|.$$

Suppose now that we redefine $(\mathbf{G}_0, \mathbf{G})$ to be an orthogonal matrix in the $\widehat{\boldsymbol{\Sigma}}$ inner product, $(\mathbf{G}_0, \mathbf{G})^T \times \widehat{\boldsymbol{\Sigma}}(\mathbf{G}_0, \mathbf{G}) = I_p$. Then $\mathbf{G}_0^T \widehat{\boldsymbol{\Sigma}} \mathbf{G}_0 = I_{p-d}$ and, letting $\mathbf{G}^* = \widehat{\boldsymbol{\Sigma}}^{1/2} \mathbf{G}$, (14) can be written as a function of $\mathbf{G}^*$ with $\mathbf{G}^{*T} \mathbf{G}^* = I_d$,

$$L_{\text{PFC}}(\mathbf{G}^*) = -(n/2)\log|\mathbf{G}^{*T}\{I_p - \widehat{\boldsymbol{\Sigma}}_{\text{sir}}\}\mathbf{G}^*|.$$

This objective function results in the standard SIR estimates since it is maximized by setting the columns of $\mathbf{G}^*$ to be the matrix $\widehat{\mathbf{G}}^*$ whose columns are the first $d$ eigenvectors of $\widehat{\boldsymbol{\Sigma}}_{\text{sir}}$. This solution is then back-transformed to the original $\mathbf{X}$ scale so that $\widehat{\mathcal{S}}_{\boldsymbol{\Gamma}} = \widehat{\boldsymbol{\Sigma}}^{-1/2} \text{span}(\widehat{\mathbf{G}}^*)$. From this we see that, starting from the normal likelihood for (13), the SIR solution is found by using a sample-based inner product. Relative to the likelihood, this can have a costly effect of neglecting the information in $\widehat{\boldsymbol{\Sigma}}$ when estimating the sufficient reduction.

While SIR does not require normality or a particular structure for $\text{Var}(\mathbf{X}_y)$, it is known that its operating characteristics can vary widely depending on these features. Bura and Cook (2001) argued that generally SIR performs the best under normality and that its performance can degrade when $\text{Var}(\mathbf{X}_y)$ is not constant. Cook and Ni (2005) showed that SIR can be very inefficient when $\text{Var}(\mathbf{X}_y)$ is not constant and they provided a new model-free method called inverse regression estimation (IRE) that can dominate SIR in applications. From these and other articles, we would expect SIR to be at its best under the models considered here, which all have both normality and constant $\text{Var}(\mathbf{X}_y)$.

Like SIR, the fundamental population characteristics of the PC and PFC methods considered here do not hinge on normality. From (14) and Proposition 4, the normalized partially maximized log likelihood $L_{\text{PFC}}(\mathbf{G})/n$ converges in probability to

$$\widetilde{L}_{\text{PFC}}(\mathbf{G}) = -(1/2)\log|\mathbf{G}_0^T \boldsymbol{\Sigma} \mathbf{G}_0|$$
$$- (1/2)\log|\mathbf{G}^T \boldsymbol{\Sigma}_{\text{res}} \mathbf{G}|.$$

We then have:

PROPOSITION 5. *Assume the conditions of Proposition 4. Then* $\boldsymbol{\Gamma} = \arg\max_{\mathbf{G}} \widetilde{L}_{\text{PFC}}(\mathbf{G})$.

This proposition says that the likelihood objective function arising from the extended PFC model (13) produces Fisher consistent estimates when the errors are uncorrelated, but not necessarily normal, suggesting that normality *per se* is not crucial for the type of analysis suggested in this article.

### 6.4 Illustration via Simulation

A simulation study was conducted to illustrate some of the results to this point. We generated $Y$ as a $N(0, \sigma_Y^2)$ random variable, and then generated $\mathbf{X}_y$ according to model (12) with $p = 10$, sample size $n = 250$ and $\boldsymbol{\Gamma} = (1, 0, \ldots, 0)^T$. As in previous simulations, $\dim(\mathcal{S}_{\boldsymbol{\Gamma}}) = 1$ to allow straightforward comparisons with OLS. In reference to model (13), the data were generated with $\boldsymbol{\Omega}_0 = \sigma_0 I_{p-1}$ and $\boldsymbol{\Omega} = \sigma$. Four estimators were applied to each data set: OLS, SIR with eight slices and PFC$_{\text{PC}}$ based on the likelihood for the extended PFC model (13) using the slicing construction (4) with eight slices for $\mathbf{f}_y$. To expand the comparisons, we also included a recent semiparametric estimator RMAVE (Xia et al., 2002), which is based on local linear smoothing of the forward regression with adaptive weights and, like SIR, is expected to do well in the context of this simulation. The results were summarized by computing the angle between each of the four estimates and $\mathcal{S}_{\boldsymbol{\Gamma}}$.

The angles plotted in Figure 2 are averages taken over 500 replications. Figure 2(a) is a plot of the average angle versus the error standard deviation $\sigma$ for the signal, with $\sigma_Y = \sigma_0 = 1$. Clearly, the likelihood-based estimator PFC$_{\text{PC}}$ dominates SIR, OLS and RMAVE, except when $\sigma$ is close to 1, so the three variances in the simulation are close. The RMAVE estimator is indistinguishable from OLS in all of the simulations of Figure 2. In Figure 2(b) $\sigma_Y$ was varied while holding $\sigma = \sigma_0 = 1$. Again we see that PFC$_{\text{PC}}$ dominates over most of the plot.

The error standard deviation $\sigma_0$ for the inactive predictors was varied for the construction of Figure 2(c). Here the results are of a fundamentally different character. PFC$_{PC}$ performed the best for $\sigma_0 < 1$, and the three methods are roughly equivalent for the larger values of $\sigma_0$. But in the middle region, PFC$_{PC}$ performed the worst. This poor performance arises because the estimate was computed using the PC candidate set, and in the simulations when $\sigma_0 = \sqrt{2}$ the population eigenvalues of $\boldsymbol{\Sigma} = 2I_p$ are equal. When $\boldsymbol{\Sigma}$ is spherical, the principal components are arbitrary and cannot be expected to convey any useful information. These observations allow us to guess about the qualitative behavior of the four estimators in Figure 2(a) when $\sigma < 1$ and in Figure 2(b) when $\sigma_y < 1$. For instance, as $\sigma_y$ goes to 0 in Figure 2(b), all methods should exhibit deteriorating performance, but PFC$_{PC}$ should again perform the worst since the eigenvalues of $\boldsymbol{\Sigma}$ converge to 1. These guesses are confirmed by simulations (not shown).

Figure 2(d) was constructed as Figure 2(c), except the estimator PFC$_{PC}$ was replaced by the estimator PFC$_{all}$ based on the candidate set of PC, PFC and RC directions. The PFC$_{all}$ estimate of $\mathcal{S}_{\boldsymbol{\Gamma}}$ always performed as well as or better than SIR, OLS and RMAVE, except in a neighborhood around $\sigma_0 = \sqrt{2}$. The vector that maximized the likelihood always came from the PC candidate set for $\sigma_0 \leq 0.75$, from the PFC candidate set for $1 \leq \sigma_0 \leq 1.25$, and from the residual candidate set for $\sigma_0 \geq 2$. At $\sigma_0 = 1.5$ this vector came about equally from the residual and

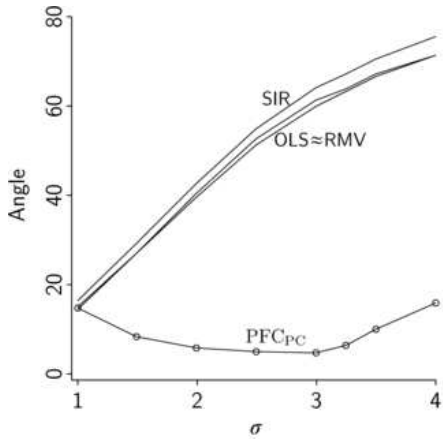
(a) Angle vs $\sigma$, with $\sigma_0 = \sigma_Y = 1$

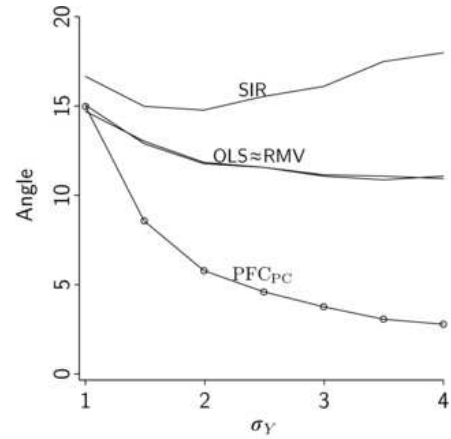
(b) Angle vs $\sigma_Y$, with $\sigma_0 = \sigma = 1$

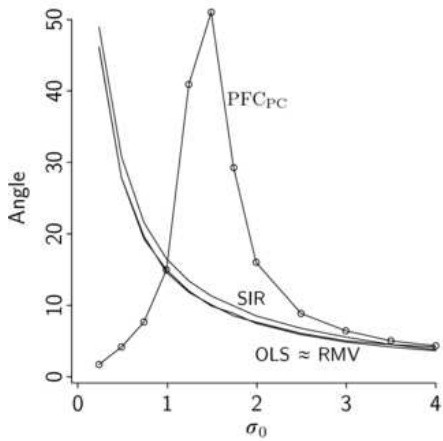
(c) Angle vs $\sigma_0$, $\sigma = \sigma_Y = 1$

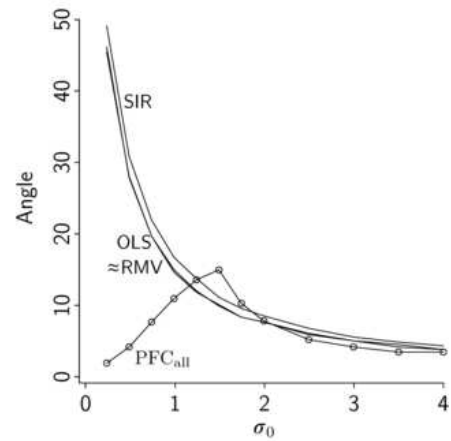
(d) Angle vs $\sigma_0$, $\sigma = \sigma_Y = 1$

FIG. 2. *Simulation results based on model (12). In* (a)–(c) *the likelihood-based estimator was computed from the PC candidate set. In* (d) *the estimator was computed from the full candidate set containing the PC, PFC and RC directions. RMV is short for RMAVE.*

PFC candidate sets. These results may provide some intuition into operational characteristics of the likelihood as reflected through the three matrices given in Proposition 4. For this simulation example those matrices are

$$\Sigma = \Gamma_0 \Gamma_0^T \sigma_0^2 + \Gamma \Gamma^T (\sigma^2 + \sigma_Y^2),$$
$$\Sigma_{\text{fit}} = \Gamma \Gamma^T \sigma_Y^2,$$
$$\Sigma_{\text{res}} = \Gamma_0 \Gamma_0^T \sigma_0^2 + \Gamma \Gamma^T \sigma^2.$$

When $\sigma_0$ is small, the first sample PC direction evidently provides the best estimate of $\mathcal{S}_\Gamma$. When $\sigma_0$ is large, we can gain information on $\mathcal{S}_\Gamma$ from the smallest PC or the largest PFC. Evidently, the error variation that comes with $\sigma_0$ causes the smallest PC to be less reliable than the largest PFC. When $\sigma_0^2 = \sigma^2 + \sigma_Y^2$, the PC directions provide no information on $\mathcal{S}_\Gamma$, but the PFC and residual directions can both provide information.

To gain insights about the potential advantages of pursuing the maximum likelihood estimator, we used a gradient algorithm for Grassmann manifolds (Edelman, Arias and Smith, 1998) to find a local maximum of the likelihood, starting with the best direction from the full candidate set. The local likelihood solution resulted in improvements all along the "PFC$_{\text{all}}$" curve, with the greatest improvement in a neighborhood around $\sigma_0 = 1.5$. For instance, at $\sigma_0 = 1.5$, the average angle for the local likelihood solution was about 11 degrees, which is quite close to the average angles for SIR and OLS, while the average angle shown in Figure 2(d) for the full candidate set is about 15 degrees. On balance, the local likelihood gave solutions that were never worse and were sometimes much better than those of the three competing methods.

SIR and RMAVE are model-free methods of dimension reduction for regression. Normally some loss would be expected relative to likelihood-based methods when the model holds. But the magnitude of the loss in this simulation, particularly in Figures 2(a) and 2(b), is surprising. The behavior of SIR is likely explained in part by the discussion in Section 6.3. Some authors (see, e.g., L. Li and H. Li, 2004; and Chiaromonte and Martinelli, 2002) have used principal components to reduce the dimension of the predictor vector prior to a second round of reduction using SIR. The results of this simulation raise questions regarding such methodology generally. If we are in a situation like Figure 2(a) or 2(b) where principal components do well, the motivation for switching to SIR seems unclear. On the other hand, if we are in a situation like Figure 2(c) with $\sigma_0 \approx \sqrt{2}$, then there seems little justification for using principal components in the first place.

### 6.5 Model Selection

The dimension $d$ of the reductive subspace was assumed known in the discussion of the extended PFC model (13), but inference on $d$, which is in effect a model selection parameter, may be required in practice. Likelihood methods are a natural first choice, including penalized log likelihood methods like AIC and BIC. In this section we briefly consider one possibility for inference on $d$, and include two simple illustrative examples.

If we set $d = p$, then we can take $\Gamma = I_p$ and the extended PFC model (13) reduces to the standard multivariate normal linear model, $\mathbf{X}_y = \boldsymbol{\mu} + \boldsymbol{\beta} \mathbf{f}_y + \boldsymbol{\Omega}\boldsymbol{\varepsilon}$, which we call the "full model." All extended PFC models with $d < p$ are properly nested within the full model and may be tested against it by using a likelihood ratio. Let $\Lambda_d$ denote $-2$ times the log likelihood ratio for comparing an extended PFC model to the full model. The dimension of the Grassmann manifold $\mathcal{G}_{p \times d}$ is $d(p-d)$, which is the number of real parameters needed to determine $\mathcal{S}_\Gamma$ and to simultaneously determine $\mathcal{S}_\Gamma^\perp$. From this it can be verified that, under the PFC model of the null hypothesis, $\Lambda_d$ is asymptotically chi-squared with $r(p-d)$ degrees of freedom.

We use the Horse Mussel data (Cook and Weisberg, 1994) for the first example. The response is the logarithm of muscle mass, and the $p = 4$ predictors are the logarithms of shell height, length, mass and width. Scatterplots of the predictor versus the response indicate that $\mathbf{f}_y = y - \bar{y}$ is a reasonable choice. The extended PFC model (13) with $d = 1$ was fitted by using Grassmann optimization with the starting value chosen as the best direction from the full candidate set. This gave $\Lambda_1 = 3.3$ with three degrees of freedom, indicating that these data provide little information to distinguish between the full model and the PFC model with $d = 1$. The pairwise sample correlations between the estimated sufficient reduction, the first PC and the first PFC were all essentially 1, so the first PC produces an equivalent solution for these data. However, PC regression by itself does not allow one to infer strong independence, $(Y, \Gamma^T \mathbf{X}) \perp\!\!\!\perp \Gamma_0^T \mathbf{X}$. Using the cubic option for $\mathbf{f}_y$ produced the same conclusions.

Fearn's (1983; see also Cook, 1998, page 175) calibration data are the basis for the second example.

The response is the protein content of a sample of ground wheat, and the predictors are $-\log(\text{reflectance})$ of NIR radiation at $p = 6$ wavelengths. The predictors are highly correlated in these data, with pairwise sample correlations ranging between 0.92 and 0.9991. As in the previous example, $\mathbf{f}_y = y - \bar{y}$ seemed to be a reasonable choice. Fitting the extended PCF model with this $\mathbf{f}_y$ gave $\Lambda_1 = 29.1$ with five degrees of freedom and $\Lambda_2 = 2.6$ with four degrees of freedom. Consequently, we infer that the sufficient reduction is composed of two linear combinations of the predictors, which can be viewed in a three-dimensional plot with the response. In contrast to the previous example, here there does not seem to be an easily described relationship between the estimated sufficient reduction and the principal components. All of the principal components are related to the sufficient reduction in varying degrees, the strongest relationships involving the second, third and sixth components.

Several data sets from the literature were studied similarly. The conclusion that $d < p$, and sometimes substantially so, was the rule rather than the exception. In the next section we consider more general versions of the PC and PFC models by allowing for unstructured errors. This will provide a closer connection with some standard methodology, and may give intuition about the common practice of standardizing the data prior to computing principal components. We concluded from this that the kinds of models proposed here will likely have wide applicability in practice.

## 7. PC AND PFC MODELS WITH UNSTRUCTURED ERRORS

Suppose that $\mathbf{X}_y$ follows the general PC model

$$(16) \qquad \mathbf{X}_y = \boldsymbol{\mu} + \boldsymbol{\Gamma}\boldsymbol{\nu}_y + \sigma\boldsymbol{\Delta}^{1/2}\boldsymbol{\varepsilon},$$

where the parameters and the error $\boldsymbol{\varepsilon}$ have the same structure as in model (2) and the conditional covariance matrix $\text{Var}(\mathbf{X}_y) = \sigma^2\boldsymbol{\Delta} > 0$. Then we have the following.

PROPOSITION 6. *Under model (16), the distribution of $Y|\mathbf{X}$ is the same as the distribution of $Y|\boldsymbol{\Gamma}^T\boldsymbol{\Delta}^{-1}\mathbf{X}$ for all values of $\mathbf{X}$.*

We first consider implications of this proposition when $\boldsymbol{\Delta}$ is known, and then turn to the case in which $\text{Var}(\mathbf{X}_y)$ is unknown.

### 7.1 $\boldsymbol{\Delta}$ known

The essential variance condition in the PC model (2) and the PFC model (5) is that $\text{Var}(\mathbf{X}_y) = \sigma^2\boldsymbol{\Delta}$, where $\boldsymbol{\Delta}$ is known but not necessarily the identity. According to Proposition 6 a sufficient reduction for (16) is $\boldsymbol{\Gamma}^T\boldsymbol{\Delta}^{-1}\mathbf{X}$. Letting $\mathbf{Z}_y = \boldsymbol{\Delta}^{-1/2}\mathbf{X}_y$, we have

$$\mathbf{Z}_y = \boldsymbol{\Delta}^{-1/2}\boldsymbol{\mu} + \boldsymbol{\Delta}^{-1/2}\boldsymbol{\Gamma}\boldsymbol{\nu}_y + \sigma\boldsymbol{\varepsilon}$$
$$= \boldsymbol{\mu}^* + \boldsymbol{\Gamma}^*\boldsymbol{\nu}_y^* + \sigma\boldsymbol{\varepsilon},$$

where the columns of $\boldsymbol{\Gamma}^*$ are an orthonormal basis for $\text{span}(\boldsymbol{\Delta}^{-1/2}\boldsymbol{\Gamma})$ and $\boldsymbol{\nu}_y^*$ is the corresponding coordinate function. It follows that $\mathcal{S}_{\boldsymbol{\Gamma}} = \boldsymbol{\Delta}^{1/2}\mathcal{S}_{\boldsymbol{\Gamma}^*}$ and thus that the coordinates of the sufficient reduction are in $\boldsymbol{\Delta}^{-1}\mathcal{S}_{\boldsymbol{\Gamma}} = \boldsymbol{\Delta}^{-1/2}\mathcal{S}_{\boldsymbol{\Gamma}^*}$. In short, the required reductive subspace $\mathcal{S}_{\boldsymbol{\Delta}^{-1}\boldsymbol{\Gamma}}$ is estimated by the span of $\boldsymbol{\Delta}^{-1/2}$ times the first $d$ eigenvectors of $\boldsymbol{\Delta}^{-1/2}\widehat{\boldsymbol{\Sigma}}\boldsymbol{\Delta}^{-1/2}$. An implication of this result is that principal components computed in the standardized $\mathbf{Z}$-scale are appropriate reductions for both $\mathbf{X}$ and $\mathbf{Z}$, because

$$\boldsymbol{\Gamma}^{*T}\mathbf{Z} = (\boldsymbol{\Gamma}^{*T}\boldsymbol{\Delta}^{-1/2})(\boldsymbol{\Delta}^{1/2}\mathbf{Z}) = (\boldsymbol{\Delta}^{-1/2}\boldsymbol{\Gamma}^*)^T\mathbf{X}.$$

Turning to the general PFC model

$$(17) \qquad \mathbf{X}_y = \boldsymbol{\mu} + \boldsymbol{\Gamma}\boldsymbol{\beta}\mathbf{f}_y + \sigma\boldsymbol{\Delta}^{1/2}\boldsymbol{\varepsilon},$$

and following the discussion of model (16), $\boldsymbol{\Gamma}^T\boldsymbol{\Delta}^{-1}\mathbf{X}$ is again a sufficient reduction. The maximum likelihood estimate of the reductive subspace is the span of $\boldsymbol{\Delta}^{-1/2}$ times the first $d$ eigenvalues of $\boldsymbol{\Delta}^{-1/2}\widehat{\boldsymbol{\Sigma}}_{\text{fit}}\boldsymbol{\Delta}^{-1/2}$.

### 7.2 Var($\mathbf{X}_y$) unknown

We now turn to the PFC model (17) with $\text{Var}(\mathbf{X}_y)$ unknown. For notational convenience, redefine $\boldsymbol{\Delta} = \text{Var}(\mathbf{X}_y)$, absorbing the scale parameter $\sigma^2$ into the definition of $\boldsymbol{\Delta}$. Maximizing the log likelihood over $\boldsymbol{\beta}$ and $\boldsymbol{\mu}$, the resulting partially maximized log likelihood

$$\begin{aligned}(18) \quad & (-n/2)\log|\mathbf{D}| \\ & - (n/2)\text{trace}[\mathbf{D}^{-1/2}\widehat{\boldsymbol{\Sigma}}\mathbf{D}^{-1/2} \\ & \qquad - P_{\mathbf{D}^{-1/2}\mathbf{G}}\mathbf{D}^{-1/2}\widehat{\boldsymbol{\Sigma}}_{\text{fit}}\mathbf{D}^{-1/2}]\end{aligned}$$

is a function of possible values $\mathbf{D}$ and $\mathbf{G}$ for $\boldsymbol{\Delta}$ and $\boldsymbol{\Gamma}$. For fixed $\mathbf{D}$ this function is maximized by choosing $\mathbf{D}^{-1/2}\mathbf{G}$ to be a basis for the span of the first $d$ eigenvectors of $\mathbf{D}^{-1/2}\widehat{\boldsymbol{\Sigma}}_{\mathrm{fit}}\mathbf{D}^{-1/2}$, yielding another partially maximized log likelihood $K(\mathbf{D})$,

$$K(\mathbf{D})/n = (-1/2)\log|\mathbf{D}|$$
$$- (1/2)\bigg\{\operatorname{trace}[\mathbf{D}^{-1/2}\widehat{\boldsymbol{\Sigma}}\mathbf{D}^{-1/2}]$$
$$- \sum_{i=1}^{d} \lambda_i(\mathbf{D}^{-1/2}\widehat{\boldsymbol{\Sigma}}_{\mathrm{fit}}\mathbf{D}^{-1/2})\bigg\}$$
$$= (-1/2)\log|\mathbf{D}|$$
$$- (1/2)\bigg\{\operatorname{trace}[\mathbf{D}^{-1}\widehat{\boldsymbol{\Sigma}}_{\mathrm{res}}]$$
$$+ \sum_{i=d+1}^{p} \lambda_i(\mathbf{D}^{-1}\widehat{\boldsymbol{\Sigma}}_{\mathrm{fit}})\bigg\},$$

where $\lambda_i(\mathbf{A})$ indicates the $i$th eigenvalue of $\mathbf{A}$, and the second equality was found by substituting $\widehat{\boldsymbol{\Sigma}} = \widehat{\boldsymbol{\Sigma}}_{\mathrm{fit}} + \widehat{\boldsymbol{\Sigma}}_{\mathrm{res}}$. The first two terms alone,

$$(-1/2)\log|\mathbf{D}| - (1/2)\operatorname{trace}[\mathbf{D}^{-1}\widehat{\boldsymbol{\Sigma}}_{\mathrm{res}}],$$

are maximized by $\widehat{\boldsymbol{\Delta}} = \widehat{\boldsymbol{\Sigma}}_{\mathrm{res}}$ and this is a consistent estimator of $\boldsymbol{\Delta}$. The final term,

$$-(1/2)\sum_{i=d+1}^{p} \lambda_i(\mathbf{D}^{-1}\widehat{\boldsymbol{\Sigma}}_{\mathrm{fit}}),$$

reflects the fact that we may have $r > d$ and thus that there is an error component in $\widehat{\boldsymbol{\Sigma}}_{\mathrm{fit}}$ due to overfitting. Since it is assumed that $\operatorname{rank}(\widehat{\boldsymbol{\Sigma}}_{\mathrm{fit}}) \geq d$, this term will not be present if $r = d$, and use of $\widehat{\boldsymbol{\Delta}} = \widehat{\boldsymbol{\Sigma}}_{\mathrm{res}}$ should be reasonable if $r$ is not much larger than $d$. However, the final term may be important if $r$ is substantially larger than $d$.

Once $\widehat{\boldsymbol{\Delta}}$ is determined, the estimate of the reductive subspace is the span of $\widehat{\boldsymbol{\Delta}}^{-1/2}$ times the first $d$ eigenvectors of $\widehat{\boldsymbol{\Delta}}^{-1/2}\widehat{\boldsymbol{\Sigma}}_{\mathrm{fit}}\widehat{\boldsymbol{\Delta}}^{-1/2}$. This is the same as the estimator in the case where $\boldsymbol{\Delta}$ is known, except $\widehat{\boldsymbol{\Delta}}$ is substituted for $\boldsymbol{\Delta}$.

### 7.3 Prior Data Standardization for PC's

Reduction by principal components is often based on the marginal correlation matrix of the predictors rather than on $\widehat{\boldsymbol{\Sigma}}$ (see, e.g., Jolliffe, 2002, page 169; L. Li and H. Li, 2004). In this section we provide a population-level discussion that may shed light on the appropriateness of this practice.

The contours of the conditional covariance matrix are spherical in the first PC model (2), $\operatorname{Var}(\mathbf{X}_y) = \sigma^2 I_p$, and the contours of $\boldsymbol{\Sigma} = \sigma^2 I_p + \boldsymbol{\Gamma}\operatorname{Var}(\boldsymbol{\nu}_Y)\boldsymbol{\Gamma}^T$ are elliptical. Generally, the method of maximum likelihood treats $\operatorname{Var}(\mathbf{X}_y)$ as a reference point, with the impact of the response being embodied in the eigenvectors of $\boldsymbol{\Sigma}$ relative to $\operatorname{Var}(\mathbf{X}_y)$. In the context of model (2), these eigenvectors are the same as the eigenvectors of $\boldsymbol{\Sigma}$. When passing from $\operatorname{Var}(\mathbf{X}_y) = \sigma^2 I_p$ to $\boldsymbol{\Sigma}$, the response distorts the conditional variance by "pulling" its spherical contours parallel to the reductive subspace, which is then spanned by the first $d$ principal components. The same ideas work for the first PFC model (5), except more information is supplied to the mean function $\operatorname{E}(\mathbf{X}_y)$, with the consequence that the marginal covariance matrix of the predictors $\boldsymbol{\Sigma}$ is replaced by the covariance matrix of the fitted values $\boldsymbol{\Sigma}_{\mathrm{fit}}$.

Likelihood estimation in all of the other normal models considered in this article can be interpreted similarly, but the calculations become more involved because the reference point $\operatorname{Var}(\mathbf{X}_y)$ is no longer spherical. Consider first the general PC model (16) with $\boldsymbol{\Delta}$ known. The reference point $\operatorname{Var}(\mathbf{X}_y) = \sigma^2 \boldsymbol{\Delta}$ no longer has spherical contours, so the eigenstructure of $\boldsymbol{\Sigma}$ is not sufficient to find the reductive subspace. To find the eigenvectors of $\boldsymbol{\Sigma}$ relative to $\operatorname{Var}(\mathbf{X}_y)$, first construct the standardized predictors $\mathbf{Z} = \boldsymbol{\Delta}^{-1/2}\mathbf{X}$. The reductive subspace in the $\mathbf{Z}$-scale is then the span of the first $d$ eigenvectors of $\operatorname{Var}(\mathbf{Z}) = \boldsymbol{\Delta}^{-1/2}\boldsymbol{\Sigma}\boldsymbol{\Delta}^{-1/2}$, because the contours of $\operatorname{Var}(\mathbf{Z}_y) = \sigma^2 I_p$ are spherical. The final step is to return to the $\mathbf{X}$-scale by multiplying these eigenvectors by $\boldsymbol{\Delta}^{-1/2}$. The general PFC model (17) follows the same pattern, as may be seen from the discussion at the end of Section 7.2.

It may now be clear why the approach in this article provides little support for the common practice of standardizing the predictors so that the covariance matrix of the new predictors $\mathbf{W} = \operatorname{diag}(\boldsymbol{\Sigma})^{-1/2}\mathbf{X}$ is a correlation matrix. The discussion of the PC model in Section 7.1 indicated that standardization should be based on $\boldsymbol{\Delta}$, not $\operatorname{diag}(\boldsymbol{\Sigma})$, followed by back transformation to the original scale. Reduction using the eigenvectors of $\operatorname{Var}(\mathbf{W})$ requires that $\operatorname{Var}(\mathbf{W}_y)$ be spherical, but this requirement will not generally be met. As a consequence, the eigenvectors of $\operatorname{Var}(\mathbf{W})$ may have little useful relation to the reductive subspace.

As a simple example, consider the simulation model (7), where the predictors are marginally and conditionally independent, with $\operatorname{Var}(\mathbf{X}_y) = \sigma^2 I_p$. Consequently, $\operatorname{Var}(\mathbf{W}) = I_p$ and the principal components of the standardized predictors do not convey useful information. However, $\text{Var}(\mathbf{W}_y)$ is not spherical, so the sufficient reduction should be computed from the eigenvectors of $\text{Var}(\mathbf{W}) = I_p$ relative to $\text{Var}(\mathbf{W}_y)$.

### 7.4 Connection with OLS for Y—X

Suppose that we adopt model (17) with $\mathbf{f}_y = (y - \bar{y})$ and $\boldsymbol{\Delta}$ unknown, again absorbing $\sigma^2$ into $\boldsymbol{\Delta}$. In this case, $d = r = 1$, $\boldsymbol{\beta}$ is a scalar and $\boldsymbol{\Gamma} \in \mathbb{R}^p$. Letting $\widehat{\mathbf{C}} = \widehat{\text{Cov}}(\mathbf{X}, Y)$,

$$\widehat{\boldsymbol{\Sigma}}_{\text{fit}} = \frac{\widehat{\mathbf{C}}\widehat{\mathbf{C}}^T}{\hat{\sigma}_Y^2},$$

which is of rank 1. Consequently, it follows from the previous section that

$$\widehat{\boldsymbol{\Delta}} = \widehat{\boldsymbol{\Sigma}}_{\text{res}} = \widehat{\boldsymbol{\Sigma}} - \frac{\widehat{\mathbf{C}}\widehat{\mathbf{C}}^T}{\hat{\sigma}_Y^2}$$

and that
$\widehat{\boldsymbol{\Sigma}}_{\text{res}}^{-1} = \widehat{\boldsymbol{\Sigma}}^{-1} + \widehat{\boldsymbol{\Sigma}}^{-1}\widehat{\mathbf{C}}[1 - \widehat{\mathbf{C}}^T\widehat{\boldsymbol{\Sigma}}^{-1}\widehat{\mathbf{C}}/\hat{\sigma}_Y^2]^{-1}\widehat{\mathbf{C}}^T\widehat{\boldsymbol{\Sigma}}^{-1}/\hat{\sigma}_Y^2$.
The estimate $\widehat{\mathcal{S}}_{\boldsymbol{\Delta}^{-1}\boldsymbol{\Gamma}}$ is the span of $\widehat{\boldsymbol{\Sigma}}_{\text{res}}^{-1/2}$ times the eigenvector corresponding to the largest eigenvalue of $\widehat{\boldsymbol{\Sigma}}_{\text{res}}^{-1/2}\widehat{\boldsymbol{\Sigma}}_{\text{fit}}\widehat{\boldsymbol{\Sigma}}_{\text{res}}^{-1/2}$. But $\widehat{\boldsymbol{\Sigma}}_{\text{fit}}$ has rank 1, and consequently the first (nonnormalized) eigenvector is $\widehat{\boldsymbol{\Sigma}}_{\text{res}}^{-1/2}\widehat{\mathbf{C}}$. Multiplying this by $\widehat{\boldsymbol{\Sigma}}_{\text{res}}^{-1/2}$, we have that $\widehat{\mathcal{S}}_{\boldsymbol{\Delta}^{-1}\boldsymbol{\Gamma}} = \text{span}(\widehat{\boldsymbol{\Sigma}}_{\text{res}}^{-1}\widehat{\mathbf{C}})$. Next, $\widehat{\boldsymbol{\Sigma}}_{\text{res}}^{-1}\widehat{\mathbf{C}} = \widehat{\boldsymbol{\Sigma}}^{-1}\widehat{\mathbf{C}} \times$ constant, so the inverse method yields OLS under model (17) with $\mathbf{f}_y = (y - \bar{y})$: $\widehat{\mathcal{S}}_{\boldsymbol{\Delta}^{-1}\boldsymbol{\Gamma}} = \text{span}(\widehat{\boldsymbol{\alpha}})$. The constant above is $\widehat{\mathbf{C}}^T\widehat{\boldsymbol{\Sigma}}^{-1}\widehat{\mathbf{C}}/(\hat{\sigma}_Y^2 - \widehat{\mathbf{C}}^T\widehat{\boldsymbol{\Sigma}}^{-1}\widehat{\mathbf{C}})$. This connection with OLS requires that $d = r = 1$ and that $\mathbf{f}_y = (y - \bar{y})$. Estimators based on (17) and other values of $d$ and $r$ may thus be regarded as a subspace generalization of OLS.

The relation $\widehat{\mathcal{S}}_{\boldsymbol{\Delta}^{-1}\boldsymbol{\Gamma}} = \text{span}(\widehat{\boldsymbol{\alpha}})$ that holds under model (17) allows us to reinterpret the results for OLS, PC and PFC in Figures 1 and 2 as a comparison between three estimators based on inverse models with different structures for $\text{E}(\mathbf{X}_y)$ and $\text{Var}(\mathbf{X}_y)$. The relative performance of the PC (2) and PFC (5) estimators in Figure 1 suggests that substantial gains are possible by modeling the inverse mean function. On the other hand, the relative performance of OLS (17) and the extended PFC estimator (13) in Figure 2 indicates that there are substantial costs associated with estimating $\boldsymbol{\Delta}$. These conclusions point to the extended PFC model (13) as a particularly useful target for dimension reduction in practice, since it requires a model for the inverse mean and avoids estimation of $\boldsymbol{\Delta}$ when appropriate.

### 7.5 Connection with SIR

As in Section 6.3, we must use the slice basis function (4) to relate the SIR estimator of the reductive subspace $\mathcal{S}_{\boldsymbol{\Delta}^{-1}\boldsymbol{\Gamma}}$ to the estimator obtained from model (17) using $\widehat{\boldsymbol{\Delta}} = \widehat{\boldsymbol{\Sigma}}_{\text{res}}$ as the estimator of $\boldsymbol{\Delta}$. Letting $\hat{\ell}_i$ denote an eigenvector of the normalized SIR matrix (cf. Section 6.3)

$$\widehat{\boldsymbol{\Sigma}}_{\text{sir}} = \widehat{\boldsymbol{\Sigma}}^{-1/2}\widehat{\boldsymbol{\Sigma}}_{\text{fit}}\widehat{\boldsymbol{\Sigma}}^{-1/2},$$

$\widehat{\boldsymbol{\Sigma}}^{-1/2}\hat{\ell}_i$ is a non-normalized eigenvector of $\widehat{\boldsymbol{\Sigma}}^{-1}\widehat{\boldsymbol{\Sigma}}_{\text{fit}}$ with the same eigenvalues as $\widehat{\boldsymbol{\Sigma}}_{\text{sir}}$. The subspace spanned by the first $d$ eigenvectors of $\widehat{\boldsymbol{\Sigma}}^{-1}\widehat{\boldsymbol{\Sigma}}_{\text{fit}}$ gives the SIR estimate of $\mathcal{S}_{\boldsymbol{\Delta}^{-1}\boldsymbol{\Gamma}}$. Similarly, the subspace spanned by the first $d$ eigenvectors of $\widehat{\boldsymbol{\Sigma}}_{\text{res}}^{-1}\widehat{\boldsymbol{\Sigma}}_{\text{fit}}$ is the estimate of the reductive subspace from model (17), still using $\widehat{\boldsymbol{\Delta}} = \widehat{\boldsymbol{\Sigma}}_{\text{res}}$. These estimators are identical provided $\widehat{\boldsymbol{\Sigma}}_{\text{res}} > 0$, because then the eigenvectors of $\widehat{\boldsymbol{\Sigma}}^{-1}\widehat{\boldsymbol{\Sigma}}_{\text{fit}}$ and $\widehat{\boldsymbol{\Sigma}}_{\text{res}}^{-1}\widehat{\boldsymbol{\Sigma}}_{\text{fit}}$ are identical, with corresponding eigenvalues $\lambda_i$ and $\lambda_i/(1 - \lambda_i)$ (see Appendix A.7).

### 7.6 Simulations with Unstructured Errors

To help fix ideas and provide results to direct further discussion, consider data simulated from the model

(19) $$\mathbf{X}_y = \boldsymbol{\Gamma} y + \boldsymbol{\Delta}^{1/2}\boldsymbol{\varepsilon},$$

where $p = 10$, $Y$ is a normal random variable with mean 0 and standard deviation $\sigma_Y = 15$, $\boldsymbol{\Gamma} = (1, \ldots, 1)^T/\sqrt{10}$, and $\boldsymbol{\Delta}$ was generated once as $\boldsymbol{\Delta} = \mathbf{A}^T\mathbf{A}$, where $\mathbf{A}$ is a $p \times p$ matrix of independent standard normal random variables, yielding predictor variances of about 10 and correlations ranging between 0.75 and $-0.67$. Four estimators of the sufficient reduction subspace $\mathcal{S}_{\boldsymbol{\Delta}^{-1}\boldsymbol{\Gamma}}$ were computed for each data set generated in this way:

1. The OLS estimator. This is the same as the PFC estimator with $\mathbf{f}_y = (y - \bar{y})$ (cf. Section 7.4).
2. The PFC estimator with a third-degree polynomial in $y$ for $\mathbf{f}_y$, designated PFC-Poly in later plots.
3. The SIR estimator with eight slices. This is the same as the PFC estimator with the slicing option for $\mathbf{f}_y$ and $\widehat{\boldsymbol{\Delta}} = \widehat{\boldsymbol{\Sigma}}_{\text{res}}$ (cf. Section 7.5).
4. The PFC estimator using the slicing construction for $\mathbf{f}_y$ with eight slices and the true $\boldsymbol{\Delta}$, designated PFC-$\boldsymbol{\Delta}$ in later plots (cf. Section 7.2).

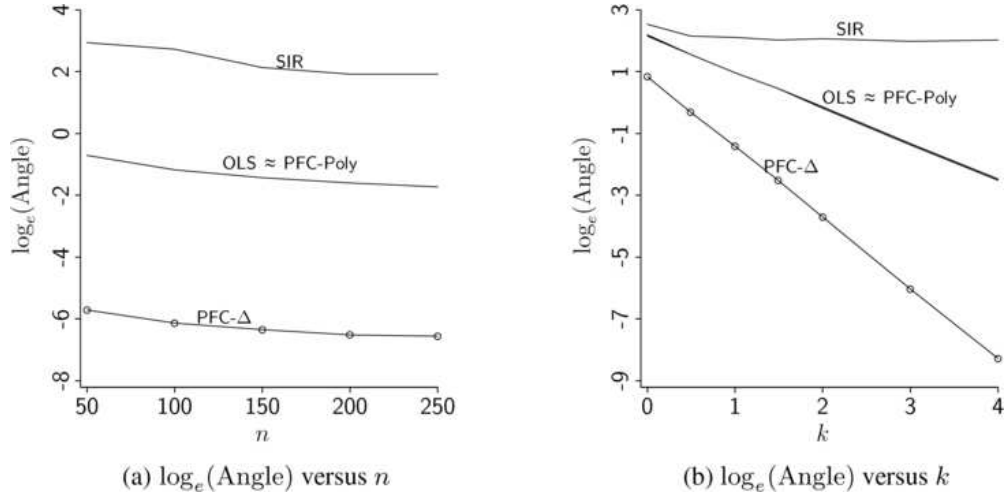

FIG. 3. *Natural logarithm of the average angle versus* (a) *sample size and* (b) *k for four estimators based on simulation model* (19) *with different covariance matrices* $\boldsymbol{\Delta}$.

Shown in Figure 3(a) are the natural logarithms of the average angles from 500 replications of each sampling configuration. The logarithms were necessary since the angles varied over several orders of magnitude. The OLS and PFC-Poly estimators were essentially indistinguishable and are represented by a single curve in Figure 3(a). These estimators seemed to do quite well, with the average angle varying between 0.51 for $n=50$ and 0.18 for $n=250$. The performance of the PFC estimator using the true $\boldsymbol{\Delta}$ was exceptional, the average angle varying between 0.0034 and 0.0014. SIR performed the worst, its average angle varying between 19.2 and 8 degrees.

PFC did exceptionally well when using the true $\boldsymbol{\Delta}$, while SIR's performance was considerably worse. The reason for this difference seems to be that SIR's slicing estimate of $\boldsymbol{\Delta}$ is biased. In the context of this example, $\boldsymbol{\Delta} = \boldsymbol{\Sigma} - \boldsymbol{\Gamma}\boldsymbol{\Gamma}^T \sigma_Y^2$. But when using the slicing construction for $\mathbf{f}_y$, $\widehat{\boldsymbol{\Delta}} = \widehat{\boldsymbol{\Sigma}}_{\text{res}}$ is a consistent estimator of

$$\boldsymbol{\Delta}_{\text{sir}} = \boldsymbol{\Sigma} - \text{Var}(\text{E}(\mathbf{X}|Y \in H_k))$$
$$= \boldsymbol{\Sigma} - \boldsymbol{\Gamma}\boldsymbol{\Gamma}^T \text{Var}(\text{E}(Y|Y \in H_k)),$$

where $H_k$ indicates slice $k$, as defined in Section 4. The difference between these two population covariance matrices is

$$\boldsymbol{\Delta} - \boldsymbol{\Delta}_{\text{sir}} = \boldsymbol{\Gamma}\boldsymbol{\Gamma}^T(\sigma_Y^2 - \text{Var}(\text{E}(Y|Y \in H_k)))$$
$$= \boldsymbol{\Gamma}\boldsymbol{\Gamma}^T \text{E}(\text{Var}(Y|Y \in H_k)),$$

which is nonzero for continuous responses. One consequence of this bias is that SIR may not be able to find an exact or near exact fit. An exact fit occurs in the context of model (17) if $\boldsymbol{\Gamma}^T \boldsymbol{\Delta} = 0$, so $\boldsymbol{\Gamma}^T \mathbf{X}$ is a deterministic function of $y$. To illustrate this phenomenon we generated data from simulation model (19) with

$$\boldsymbol{\Delta} = (cI_{10} - \boldsymbol{\Gamma}\boldsymbol{\Gamma}^T)\sigma_Y^2$$

and $c > 1$. Here $\boldsymbol{\Sigma} = c\sigma_Y^2 I_{10}$ and an exact fit occurs if $c = 1$. Values of $c < 1$ are not allowed since then $\boldsymbol{\Delta}$ will not be positive definite. With $c = 1 + 0.1/10^k$, Figure 3(b) shows the natural logarithm of the average simulation angle versus $k$ for the four estimators used in Figure 3(a) with $n = 50$. Clearly, SIR's response to increasing $k$ is negligible. At $k = 4$ the average angles for SIR, OLS and PFC-$\boldsymbol{\Delta}$ were about 7.7, 0.085 and 0.0002, respectively.

The general conclusions here are that (1) there can be a substantial cost associated with estimation of $\boldsymbol{\Delta}$ (cf. Section 7.4) and (2) the slicing construction for $\mathbf{f}_y$ may impose inherent limitations on the analysis under model (17). The first conclusion does not occur for any of the other inverse models discussed in this article, because for them $\mathcal{S}_{\boldsymbol{\Gamma}}$ is the reductive subspace, which does not require a direct estimate of $\boldsymbol{\Delta}$.

## 8. DISCUSSION

### 8.1 General Remarks

*Conditioning.* Fisher believed that if a statistic is ancillary, then inferences should be made from the conditional distribution of the data given that statistic. As a consequence of this logic, many of us have been taught and still practice what has become effectively a first principle of parametric regression: Inference should be conditioned on the observed values of the predictors, even if $Y$ and $\mathbf{X}$ are both random. It may be, however, that this principle has forced a myopic view of regression methodology. The inverse models studied in the previous sections describe the conditional distribution of $\mathbf{X}$ given $Y$ and thereby make explicit use of the randomness in $\mathbf{X}$. With these inverse models, we were able to achieve results that are superior to those from standard methods, and to those from recent dimension reduction methods. The success of these inverse models depends in part on imposing an appropriately restrictive structure on the conditional variances $\text{Var}(\mathbf{X}_y)$. They were presented, not as models for shotgun application, but as illustrations of potential, recalling Fisher's view that model specification is a matter for the practical statistician. In contrast, if we condition on the observed values of the predictors, they become known constants and the possibilities of inferring about their variance structure or utilizing prior knowledge are lost to us.

*Reductive subspaces.* The reductive subspace provides a connection between forward and inverse regressions. Starting with the PC model (2) and passing through a series of extensions, the last two PFC models considered were the extended PFC model (13), $\mathbf{X}_y = \boldsymbol{\mu} + \boldsymbol{\Gamma}\boldsymbol{\beta}\mathbf{f}_y + \boldsymbol{\Gamma}_0\boldsymbol{\Omega}_0\boldsymbol{\varepsilon}_0 + \boldsymbol{\Gamma}\boldsymbol{\Omega}\boldsymbol{\varepsilon}$, and the general PFC model (17), $\mathbf{X}_y = \boldsymbol{\mu} + \boldsymbol{\Gamma}\boldsymbol{\beta}\mathbf{f}_y + \boldsymbol{\Delta}^{1/2}\boldsymbol{\varepsilon}$, in which $\boldsymbol{\Delta}$ is unstructured and unknown. This final model connects inverse and forward regression methodology, since it is here that certain forward and inverse estimates of $\mathcal{S}_{\boldsymbol{\Delta}^{-1}\boldsymbol{\Gamma}}$ are the same. While the error structure in model (13) is restrictive, it may be useful in some applications. Perhaps more importantly, we should be aware of the possibility to develop models "between" (13) and (17) that allow us to infer simultaneously about the reductive subspace and about the relevant structure of $\text{Var}(\mathbf{X}_y)$.

*Simulation practices.* Perhaps due in part to the conditioning tradition, it seems quite common to generate $\mathbf{X}$ as a $N(0, I)$ random vector in simulation studies to compare regression methods. This practice may place notable limitations on the results of the simulation, since it implies that there is no useful information in $\text{Var}(\mathbf{X})$, as in the simulations with $\sigma_0 = \sqrt{2}$ in Figure 2(c). This will give a clear edge to some forward methods. However, as demonstrated in this article, when $(Y, \mathbf{X})$ has a joint distribution there may well be useful information in $\text{Var}(\mathbf{X})$.

*Collinearity.* The rationale for employing principal components in regression has been rather uneven. Tied closely to the presence of collinearity, reduction by principal components has been seen as a way to compensate for variance inflation in the estimates of the regression coefficients. However, collinearity played no essential role in this article, suggesting that the utility of principal component reduction is broader than previously seen.

*$n < p$.* Dimension reduction seems particularly important in regressions where "$n < p$." Many available methods encounter problems at an operational level because of the need to compute $\widehat{\boldsymbol{\Sigma}}^{-1}$. However, with the exception of Section 7, the methods described in this article do not require the computation of an inverse, and may therefore have value in regressions where $n$ is not large relative to $p$. First simulation results sustain this conjecture, particularly when the methods are used in conjunction with predictor screening at the outset. As mentioned at the end of Section 2, past studies have based predictor screening on the univariate forward regressions of $Y$ on $X_j$. However, the results here suggest that predictor screening be based on univariate inverse regressions of $X_j$ on $\mathbf{f}_y$, $j = 1, \ldots, p$.

### 8.2 Model-Based Sufficient Dimension Reduction

At the outset of his book *Statistical Methods for Research Workers*, Fisher (1941, page 1) offered the following definition:

> Statistics may be regarded as (i) the study of *populations*, (ii) as the study of *variation*, (iii) as the study of methods of the *reduction of data*.

In this statement and in his commentary that follows, Fisher seems to suggest that "reduction of data" may encompass more than just sufficient statistics; for instance, efficient statistics may be adequate for such purposes. For this reason I imagine that Fisher would not have objected to the notion of a sufficient reduction as defined in this article.

The sufficient reductions described in the previous sections are special cases of a general reductive paradigm that emerges from the following definition: A reduction $R: \mathbb{R}^p \to \mathbb{R}^q$, $q \leq p$, is *sufficient* if at least one of the following three statements holds:

*Reductive forms.*

(i) $\mathbf{X}|(Y, R(\mathbf{X})) \sim \mathbf{X}|R(\mathbf{X})$,
(ii) $Y|\mathbf{X} \sim Y|R(\mathbf{X})$,
(iii) $Y \perp\!\!\!\perp \mathbf{X}|R(\mathbf{X})$.

Statement (i) corresponds to inverse regression and requires only the conditional distribution of $\mathbf{X}|Y$. For instance, if $Y$ is a categorical response indicating one of two populations and $\mathbf{X}|Y$ is normal with mean $\boldsymbol{\mu}_y$ and covariance matrix $\boldsymbol{\Delta}$, then Fisher's linear discriminant function is a sufficient reduction (Rao, 1962). Statement (ii) corresponds to forward regression and requires only the conditional distribution of $Y|\mathbf{X}$, while statement (iii) requires the joint distribution of $(Y, \mathbf{X})$. The key point for present purposes is that these three forms are equivalent if $Y$ and $\mathbf{X}$ are both random. For example, we may determine a sufficient reduction from $\mathbf{X}|Y$ (i) and then pass that reduction to the forward regression (ii) or the joint distribution (iii) without specifying the marginal distribution of $Y$ or the conditional distribution of $Y|\mathbf{X}$. This is how all sufficient reductions were determined in the previous sections, the methods of derivation being essentially the same as the methods available for determining sufficient statistics.

The connection with sufficient statistics goes further: If we set $\mathbf{X}$ equal to the data, $\mathbf{X} = D$, and set $Y$ equal to the parameters, $Y = \theta$, then statement (i) becomes $D|(\theta, R) \sim D|R$ and we are led back to Fisher's concept of sufficiency (1). In this way, the notion of a sufficient reduction encompasses sufficient statistics as well.

While the reductions discussed in the previous sections are linear functions of $\mathbf{X}$, sufficient reductions do not have to be linear. If $\mathbf{X}_y$ is normally distributed with mean 0 and $\text{Var}(\mathbf{X}_y) = I_p + \nu_y \boldsymbol{\Gamma}\boldsymbol{\Gamma}^T$, $\nu_y > -1$, $\boldsymbol{\Gamma} \in \mathbb{R}^p$, then $(\boldsymbol{\Gamma}^T \mathbf{X})^2$ is a sufficient reduction.

### 8.3 Model-Free Sufficient Dimension Reduction

In this article I have adopted a largely Fisherian perspective: (1) Find an adequate solution to the problem of specification for the conditional distribution of $\mathbf{X}|Y$, (2) use the inverse model for $\mathbf{X}|Y$ to estimate a sufficient reduction $R(\mathbf{X})$, and then, as described in Section 8.2, (3) pass the estimated reduction to the forward regression. Lacking an inverse model, these ideas are not directly applicable because then there is no probability structure with which to determine a sufficient reduction. However, progress is still possible by restricting the search for a sufficient reduction to a specific functional form. In view of their prevalence in the previous sections, linear reductions form a natural and potentially useful class.

Consider then the reductive forms of Section 8.2 with $R(\mathbf{X}) = \mathbf{G}^T \mathbf{X}$, $\mathbf{G} \in \mathbb{R}^{p \times k}$, $k \leq p$. One linear reduction always exists because statements (i)–(iii) are trivially true with $\mathbf{G} = I_p$. If $\mathbf{G}^T \mathbf{X}$ is a linear reduction, then so is $\mathbf{A}^T \mathbf{G}^T \mathbf{X}$ for any full rank $\mathbf{A} \in \mathbb{R}^{k \times k}$, suggesting again that interests center on the *dimension reduction (DR) subspace* $\mathcal{S}_{\mathbf{G}}$. If $\mathcal{S}_{\mathbf{G}}$ is a DR subspace and $\mathcal{S}_{\mathbf{G}} \subseteq \mathcal{S}$, then $\mathcal{S}$ is also a DR subspace. There may be infinitely many DR subspaces, and it therefore becomes necessary to consider the "smallest" subspace.

There are at least two ways to define a smallest DR subspace. One way is to require a subspace $\mathcal{S}_{\min} = \min_k \mathcal{S}_{\mathbf{G}}$ with the smallest dimension. However, such subspaces are not necessarily unique and, even if they were, they do not impose sufficient structure on the regression for progress in theory or uncomplicated application in practice. Another way is to restrict attention to the class of regressions in which the intersection $\mathcal{S}_{Y|\mathbf{X}} = \bigcap \mathcal{S}_{\mathbf{G}}$ of all DR subspaces is itself a DR subspace. The *central subspace* $\mathcal{S}_{Y|\mathbf{X}}$ (Cook, 1994, 1998) then becomes the parameter of interest. The reductive subspaces $\mathcal{S}_{\boldsymbol{\Gamma}}$ and $\mathcal{S}_{\boldsymbol{\Delta}^{-1}\boldsymbol{\Gamma}}$ encountered in previous sections are instances of model-based DR subspaces. Since generally there is no forward or inverse model to tie up loose ends like high-order conditional moments, it has proven quite hard to estimate the entire central subspace without some restrictions on the regression. Nevertheless, there are successful methods that can provide useful estimates of $\mathcal{S}_{Y|\mathbf{X}}$ under conditions that are weak relative to a parsimonious forward or inverse model.

An introduction to model-free sufficient dimension reduction via central subspaces is available from Cook (1998) and the references contained therein. See Cook and Ni (2005) for recent methodology and references to recent literature.

## APPENDIX

### A.1 Propositions 1, 3 and 6

We first demonstrate Proposition 6. Propositions 1 and 3 will then follow as special cases.

To demonstrate Proposition 6 we first show that the distribution of $\mathbf{X}|(\boldsymbol{\Gamma}^T \boldsymbol{\Delta}^{-1} \mathbf{X}, Y = y)$ is the same

as the distribution of $\mathbf{X}|\mathbf{\Gamma}^T\mathbf{\Delta}^{-1}\mathbf{X}$ for all $y$. According to model (16), $\mathbf{X}_y$ is normally distributed with mean $\boldsymbol{\mu}_y = \boldsymbol{\mu} + \mathbf{\Gamma}\boldsymbol{\nu}_y$ and constant variance. Thus $\mathbf{X}|(\mathbf{\Gamma}^T\mathbf{\Delta}^{-1}\mathbf{X}, Y = y)$ is normally distributed with constant variance and mean

$$\begin{aligned}
& E(\mathbf{X}|\mathbf{\Gamma}^T\mathbf{\Delta}^{-1}\mathbf{X}, Y = y) \\
& = \boldsymbol{\mu}_y + P^T_{\mathbf{\Delta}^{-1}\mathbf{\Gamma}(\mathbf{\Delta})}(\mathbf{X} - \boldsymbol{\mu}_y) \\
& = (I_p - P^T_{\mathbf{\Delta}^{-1}\mathbf{\Gamma}(\mathbf{\Delta})})\boldsymbol{\mu} + P^T_{\mathbf{\Delta}^{-1}\mathbf{\Gamma}(\mathbf{\Delta})}\mathbf{X} \\
& \quad + (I_p - P^T_{\mathbf{\Delta}^{-1}\mathbf{\Gamma}(\mathbf{\Delta})})\mathbf{\Gamma}\boldsymbol{\nu}_y \\
& = (I_p - P^T_{\mathbf{\Delta}^{-1}\mathbf{\Gamma}(\mathbf{\Delta})})\boldsymbol{\mu} + P^T_{\mathbf{\Delta}^{-1}\mathbf{\Gamma}(\mathbf{\Delta})}\mathbf{X},
\end{aligned}$$

where $P_{\mathbf{\Delta}^{-1}\mathbf{\Gamma}(\mathbf{\Delta})}$ is the operator that projects onto $\text{span}(\mathbf{\Delta}^{-1}\mathbf{\Gamma})$ in the $\mathbf{\Delta}$ inner product. From this the last term in the second equation is 0. Since $\mathbf{X}|(\mathbf{\Gamma}^T\mathbf{\Delta}^{-1}\mathbf{X}, Y = y)$ is normally distributed with mean and variance that do not depend on $y$, it follows that the distribution of $\mathbf{X}|(\mathbf{\Gamma}^T\mathbf{\Delta}^{-1}\mathbf{X}, Y = y)$ is the same as the distribution of $\mathbf{X}|\mathbf{\Gamma}^T\mathbf{\Delta}^{-1}\mathbf{X}$ for all $y$. Consequently, $Y \perp\!\!\!\perp \mathbf{X}|\mathbf{\Gamma}^T\mathbf{\Delta}^{-1}\mathbf{X}$, which implies that $Y|\mathbf{X}$ and $Y|\mathbf{\Gamma}^T\mathbf{\Delta}^{-1}\mathbf{X}$ have identical distributions.

Proposition 1 follows immediately because $\mathbf{\Delta} = I_p$. Under Proposition 3,

$$[\text{Var}(\mathbf{X}_y)]^{-1} = \mathbf{\Gamma}_0\mathbf{\Omega}_0^{-2}\mathbf{\Gamma}_0^T + \mathbf{\Gamma}\mathbf{\Omega}^{-2}\mathbf{\Gamma}^T.$$

Consequently, $[\text{Var}(\mathbf{X}_y)]^{-1}\mathbf{\Gamma} = \mathbf{\Gamma}\mathbf{\Omega}^{-2}$ and Proposition 3 follows.

### A.2 Proposition 2

We use a different approach to demonstrate this proposition. Suppose that under model (3) the joint density or mass function $f(\mathbf{x}|y)$ of $\mathbf{X}|(Y = y)$ can be written as

$$f(\mathbf{x}|y) = h(\mathbf{x})g(\mathbf{\Gamma}^T\mathbf{x}, y),$$

where $h$ is a function that does not depend on $y$ and $g$ is a function that depends on $\mathbf{x}$ only through $\mathbf{\Gamma}^T\mathbf{x}$. It would then follow from the usual factorization theorem for sufficiency that the distribution of $\mathbf{X}|(\mathbf{\Gamma}^T\mathbf{X}, Y = y)$ is the same as the distribution of $\mathbf{X}|\mathbf{\Gamma}^T\mathbf{X}$ for all $y$, and thus $Y \perp\!\!\!\perp \mathbf{X}|\mathbf{\Gamma}^T\mathbf{X}$.

To demonstrate the required factorization, let $x_j$ be the $j$th element of $\mathbf{x}$ and, using the conditional independence of the predictors, write

$$f(\mathbf{x}|y) = \prod_{j=1}^p a_j(\eta_{yj})b_j(x_j)\exp\{x_j\eta_{yj}\}$$

$$\begin{aligned}
& = \prod_{j=1}^p a_j(\eta_{yj})b_j(x_j)\exp\{x_j(\mu_j + \boldsymbol{\gamma}_j^T\boldsymbol{\nu}_y)\} \\
& = \left[\prod_{j=1}^p b_j(x_j)\exp(x_j\mu_j)\right] \\
& \quad \times \left[\exp\{\boldsymbol{\nu}_y^T\mathbf{\Gamma}^T\mathbf{x}\}\prod_{j=1}^p a_j(\eta_{yj})\right] \\
& = h(\mathbf{x})g(\mathbf{\Gamma}^T\mathbf{x}, y).
\end{aligned}$$

### A.3 Equation (6)

Substituting $\hat{\boldsymbol{\mu}} = \bar{\mathbf{X}}$ and $\hat{\boldsymbol{\beta}}^T = (\mathbf{F}^T\mathbf{F})^{-1}\mathbf{F}^T\mathbb{X}\mathbf{G}$ into the log likelihood, we need to maximize

$$\begin{aligned}
M_{\text{PFC}} & = (-np/2)\log(s^2) \\
& \quad - (1/2s^2)\sum_y \|\mathbf{X}_y - \bar{\mathbf{X}} - P_{\mathbf{G}}\mathbb{X}^T\mathbf{F}(\mathbf{F}^T\mathbf{F})^{-1}\mathbf{f}_y\|^2 \\
& = (-np/2)\log(s^2) \\
& \quad - (1/2s^2)\bigg\{\text{trace}\bigg[\sum_y(\mathbf{X}_y - \bar{\mathbf{X}})^T(\mathbf{X}_y - \bar{\mathbf{X}})\bigg] \\
& \quad - \text{trace}\bigg[P_{\mathbf{G}}\mathbb{X}^T\mathbf{F}(\mathbf{F}^T\mathbf{F})^{-1} \\
& \quad \quad \times \sum_y \mathbf{f}_y(\mathbf{X}_y - \bar{\mathbf{X}})^T\bigg] \\
& \quad - \text{trace}\bigg[\sum_y(\mathbf{X}_y - \bar{\mathbf{X}}) \\
& \quad \quad \times \mathbf{f}_y^T(\mathbf{F}^T\mathbf{F})^{-1}\mathbf{F}^T\mathbb{X}P_{\mathbf{G}}\bigg] \\
& \quad + \text{trace}\bigg[(\mathbf{F}^T\mathbf{F})^{-1}\mathbf{F}^T\mathbb{X}P_{\mathbf{G}}\mathbb{X}^T \\
& \quad \quad \times \mathbf{F}(\mathbf{F}^T\mathbf{F})^{-1}\sum_y \mathbf{f}_y\mathbf{f}_y^T\bigg]\bigg\}.
\end{aligned}$$

But $\sum_y \mathbf{f}_y(\mathbf{X}_y - \bar{\mathbf{X}})^T = \mathbf{F}^T\mathbb{X}$ and $\sum_y \mathbf{f}_y\mathbf{f}_y^T = \mathbf{F}^T\mathbf{F}$. Thus

$$\begin{aligned}
M_{\text{PFC}} & = (-np/2)\log(s^2) \\
& \quad - (1/2s^2)\{\text{trace}[\mathbb{X}^T\mathbb{X}] - \text{trace}[P_{\mathbf{G}}\mathbb{X}^T P_{\mathbf{F}}\mathbb{X}] \\
& \quad \quad - \text{trace}[\mathbb{X}^T P_{\mathbf{F}}\mathbb{X}P_{\mathbf{G}}] \\
& \quad \quad + \text{trace}[\mathbb{X}^T P_{\mathbf{F}}\mathbb{X}P_{\mathbf{G}}]\} \\
& = (-np/2)\log(s^2)
\end{aligned}$$

$$-(1/2s^2)\{\operatorname{trace}[\mathbb{X}^T\mathbb{X}] - \operatorname{trace}[P_\mathbf{G}\mathbb{X}^T P_\mathbf{F}\mathbb{X}]\}$$
$$= (-np/2)\log(s^2)$$
$$- (n/2s^2)\{\operatorname{trace}[\widehat{\boldsymbol{\Sigma}}] - \operatorname{trace}[P_\mathbf{G}\widehat{\boldsymbol{\Sigma}}_{\mathrm{fit}}]\}.$$

### A.4 Proposition 4

Requiring the errors to be uncorrelated but not necessarily normal, it is known that $\widehat{\boldsymbol{\Sigma}} \xrightarrow{p} \boldsymbol{\Sigma}$, and

$$\boldsymbol{\Sigma} = \operatorname{Var}(\mathbf{X})$$
$$= \operatorname{E}(\operatorname{Var}(\mathbf{X}|Y)) + \operatorname{Var}(\operatorname{E}(\mathbf{X}|Y))$$
$$= \boldsymbol{\Gamma}_0 \boldsymbol{\Omega}_0^2 \boldsymbol{\Gamma}_0^T + \boldsymbol{\Gamma}\boldsymbol{\Omega}^2\boldsymbol{\Gamma}^T + \boldsymbol{\Gamma}\boldsymbol{\beta}\operatorname{Var}(\mathbf{f}_Y)\boldsymbol{\beta}^T\boldsymbol{\Gamma}^T.$$

To find the limiting value of $\widehat{\boldsymbol{\Sigma}}_{\mathrm{fit}} = \mathbb{X}^T P_\mathbf{F} \mathbb{X}/n$, use model (13) to write

$$\mathbf{X}_y^T - \bar{\mathbf{X}}^T = (\boldsymbol{\mu}^T - \bar{\mathbf{X}}^T) + \mathbf{f}_y^T\boldsymbol{\beta}^T\boldsymbol{\Gamma}^T$$
$$+ \boldsymbol{\varepsilon}_0^T\boldsymbol{\Omega}_0^T\boldsymbol{\Gamma}_0^T + \boldsymbol{\varepsilon}^T\boldsymbol{\Omega}^T\boldsymbol{\Gamma}^T$$

and thus

$$\mathbb{X} = 1_n(\boldsymbol{\mu}^T - \bar{\mathbf{X}}^T) + \mathbf{F}\boldsymbol{\beta}^T\boldsymbol{\Gamma}^T + \mathbf{R}_0\boldsymbol{\Omega}_0\boldsymbol{\Gamma}_0^T + \mathbf{R}\boldsymbol{\Omega}\boldsymbol{\Gamma}^T,$$
$$P_\mathbf{F}\mathbb{X} = \mathbf{F}\boldsymbol{\beta}^T\boldsymbol{\Gamma}^T + P_\mathbf{F}\mathbf{R}_0\boldsymbol{\Omega}_0\boldsymbol{\Gamma}_0^T + P_\mathbf{F}\mathbf{R}\boldsymbol{\Omega}\boldsymbol{\Gamma}^T,$$

where $\mathbf{R}_0$ is the $n \times p - d$ matrix with rows $\boldsymbol{\varepsilon}_0^T$ and $\mathbf{R}$ is the $n \times d$ matrix with rows $\boldsymbol{\varepsilon}^T$. From this, $\mathbb{X}^T P_\mathbf{F} \mathbb{X}/n$ contains nine terms. The first of these is $\boldsymbol{\Gamma}\boldsymbol{\beta}(\mathbf{F}^T\mathbf{F}/n)\boldsymbol{\Gamma}^T\boldsymbol{\beta}^T \xrightarrow{p} \boldsymbol{\Gamma}\boldsymbol{\beta}\operatorname{Var}(\mathbf{f}_Y)\boldsymbol{\Gamma}^T\boldsymbol{\beta}^T$. The remaining terms involve products containing either or both of the factors $\mathbf{F}^T\mathbf{R}/n$ and $\mathbf{F}^T\mathbf{R}_0/n$. Recalling that the rows (corresponding to samples) of $\mathbf{R}$ and $\mathbf{R}_0$ are independent, these factors both converge to 0 and this property forces each of the eight remaining terms to converge to 0. For instance, the last quadratic term can be written $\boldsymbol{\Gamma}\boldsymbol{\Omega}(\mathbf{R}^T P_\mathbf{F}\mathbf{R}/n)\boldsymbol{\Omega}\boldsymbol{\Gamma}^T$, and

$$\mathbf{R}^T P_\mathbf{F} \mathbf{R}/n = (\mathbf{R}^T\mathbf{F}/n)(\mathbf{F}^T\mathbf{F}/n)^-(\mathbf{F}^T\mathbf{R}/n),$$

which converges to 0 in probability by Slutsky's theorem.

Since $\widehat{\boldsymbol{\Sigma}} = \widehat{\boldsymbol{\Sigma}}_{\mathrm{fit}} + \widehat{\boldsymbol{\Sigma}}_{\mathrm{res}}$, take the difference of the limiting values for $\widehat{\boldsymbol{\Sigma}}$ and $\widehat{\boldsymbol{\Sigma}}_{\mathrm{fit}}$ to confirm the limiting value for $\widehat{\boldsymbol{\Sigma}}_{\mathrm{res}}$.

### A.5 Equation (15)

Let $\mathbf{O} = (\mathbf{O}_1, \mathbf{O}_2)$ be a partitioned $p \times p$ orthogonal matrix. Then

$$|\widehat{\boldsymbol{\Sigma}}| = |\mathbf{O}^T\widehat{\boldsymbol{\Sigma}}\mathbf{O}| = \begin{vmatrix} \mathbf{O}_1^T\widehat{\boldsymbol{\Sigma}}\mathbf{O}_1 & \mathbf{O}_1^T\widehat{\boldsymbol{\Sigma}}\mathbf{O}_2 \\ \mathbf{O}_2^T\widehat{\boldsymbol{\Sigma}}\mathbf{O}_1 & \mathbf{O}_2^T\widehat{\boldsymbol{\Sigma}}\mathbf{O}_2 \end{vmatrix}$$
$$= |\mathbf{O}_1^T\widehat{\boldsymbol{\Sigma}}\mathbf{O}_1||\mathbf{O}_2^T\widehat{\boldsymbol{\Sigma}}\mathbf{O}_2$$
$$\quad - \mathbf{O}_2^T\widehat{\boldsymbol{\Sigma}}\mathbf{O}_1(\mathbf{O}_1^T\widehat{\boldsymbol{\Sigma}}\mathbf{O}_1)^{-1}\mathbf{O}_1^T\widehat{\boldsymbol{\Sigma}}\mathbf{O}_2|$$
$$\leq |\mathbf{O}_1^T\widehat{\boldsymbol{\Sigma}}\mathbf{O}_1||\mathbf{O}_2^T\widehat{\boldsymbol{\Sigma}}\mathbf{O}_2|.$$

### A.6 Proposition 5

Let $\mathbf{C} = \boldsymbol{\beta}\operatorname{Var}(\mathbf{f}_Y)\boldsymbol{\beta}^T$ and $\mathbf{H} = \mathbf{G}_0^T\boldsymbol{\Gamma}\mathbf{C}^{1/2}$. Then it follows from Proposition 4 that

$$-2\widetilde{L}_{\mathrm{PFC}}(\mathbf{G}) = \log|\mathbf{G}_0^T\boldsymbol{\Sigma}\mathbf{G}_0| + \log|\mathbf{G}^T\boldsymbol{\Sigma}_{\mathrm{res}}\mathbf{G}|$$
$$= \log|\mathbf{G}_0^T\boldsymbol{\Sigma}_{\mathrm{res}}\mathbf{G}_0 + \mathbf{G}_0^T\boldsymbol{\Gamma}\mathbf{C}\boldsymbol{\Gamma}^T\mathbf{G}_0|$$
$$\quad + \log|\mathbf{G}^T\boldsymbol{\Sigma}_{\mathrm{res}}\mathbf{G}|$$
$$= \log|\mathbf{G}_0^T\boldsymbol{\Sigma}_{\mathrm{res}}\mathbf{G}_0|$$
$$\quad + \log|I_d + \mathbf{H}^T(\mathbf{G}_0^T\boldsymbol{\Sigma}_{\mathrm{res}}\mathbf{G}_0)^{-1}\mathbf{H}|$$
$$\quad + \log|\mathbf{G}^T\boldsymbol{\Sigma}_{\mathrm{res}}\mathbf{G}|$$
$$> \log|\mathbf{G}_0^T\boldsymbol{\Sigma}_{\mathrm{res}}\mathbf{G}_0| + \log|\mathbf{G}^T\boldsymbol{\Sigma}_{\mathrm{res}}\mathbf{G}|$$
$$\geq \log|\boldsymbol{\Sigma}_{\mathrm{res}}|$$
$$= \log|\boldsymbol{\Gamma}\boldsymbol{\Omega}^2\boldsymbol{\Gamma}^T + \boldsymbol{\Gamma}_0\boldsymbol{\Omega}_0^2\boldsymbol{\Gamma}_0^T|$$
$$= \log|\boldsymbol{\Omega}_0^2||\boldsymbol{\Omega}^2|.$$

In addition,

$$-2\widetilde{L}_{\mathrm{PFC}}(\boldsymbol{\Gamma}) = \log|\boldsymbol{\Gamma}_0^T\boldsymbol{\Sigma}\boldsymbol{\Gamma}_0| + \log|\boldsymbol{\Gamma}^T\boldsymbol{\Sigma}_{\mathrm{res}}\boldsymbol{\Gamma}|$$
$$= \log|\boldsymbol{\Omega}_0^2||\boldsymbol{\Omega}^2|.$$

Therefore $\widetilde{L}_{\mathrm{PFC}}(\mathbf{G}) \geq L(\boldsymbol{\Gamma})$ for all $\mathbf{G}$. The minimizing argument yields a unique subspace if $\widetilde{L}_{\mathrm{PFC}}(\mathbf{G}) > \widetilde{L}_{\mathrm{PFC}}(\boldsymbol{\Gamma})$ for all $\mathbf{G}$ such that $\mathbf{G}^T\mathbf{G} = I_d$, $\dim(\mathcal{S}_\mathbf{G}) = d$ and $\mathcal{S}_\mathbf{G} \neq \mathcal{S}_{\boldsymbol{\Gamma}}$.

### A.7 Eigenvectors of $\widehat{\boldsymbol{\Sigma}}^{-1}\widehat{\boldsymbol{\Sigma}}_{\mathrm{fit}}$ and $\widehat{\boldsymbol{\Sigma}}_{\mathrm{res}}^{-1}\widehat{\boldsymbol{\Sigma}}_{\mathrm{fit}}$

$$\widehat{\boldsymbol{\Sigma}}^{-1}\widehat{\boldsymbol{\Sigma}}_{\mathrm{fit}}\ell = \lambda\ell \iff \widehat{\boldsymbol{\Sigma}}_{\mathrm{fit}}\ell = \lambda\widehat{\boldsymbol{\Sigma}}\ell$$
$$\iff \widehat{\boldsymbol{\Sigma}}_{\mathrm{fit}}\ell = \lambda\widehat{\boldsymbol{\Sigma}}_{\mathrm{res}}\ell + \lambda\widehat{\boldsymbol{\Sigma}}_{\mathrm{fit}}\ell$$
$$\iff (1-\lambda)\widehat{\boldsymbol{\Sigma}}_{\mathrm{fit}}\ell = \lambda\widehat{\boldsymbol{\Sigma}}_{\mathrm{res}}\ell$$
$$\iff \widehat{\boldsymbol{\Sigma}}_{\mathrm{res}}^{-1}\widehat{\boldsymbol{\Sigma}}_{\mathrm{fit}}\ell = (\lambda/(1-\lambda))\ell.$$

The conclusion follows because $\widehat{\boldsymbol{\Sigma}}_{\mathrm{res}} = \widehat{\boldsymbol{\Sigma}} - \widehat{\boldsymbol{\Sigma}}_{\mathrm{fit}} > 0$ and $\lambda/(1-\lambda)$ is a strictly monotonic function of $\lambda$.

## ACKNOWLEDGMENTS

I benefited from discussions of this work with several friends and colleagues, including Jim Berger, Francesca Chiaromonte, Liliana Forzani, Bing Li, Lexin Li, Liqiang Ni and Sandy Weisberg. The referees provided many helpful comments.

Research for this article was supported in part by NSF Grant DMS-04-05360.